\documentclass[fleqn,usenatbib]{mnras}

\usepackage{newtxtext,newtxmath}
\usepackage{svg}
\usepackage{ulem}
\usepackage{color}
\usepackage{soul}
\usepackage{cancel}
\usepackage{import}
\usepackage{subfig}
\usepackage{amsmath}
\usepackage{makecell}
\usepackage{array}

\usepackage[T1]{fontenc}

\DeclareRobustCommand{\VAN}[3]{#2}
\let\VANthebibliography\thebibliography
\def\thebibliography{\DeclareRobustCommand{\VAN}[3]{##3}\VANthebibliography}

\usepackage{graphicx}	
\usepackage{amsmath}	

\title[photo-z from CSST flux with Random Forest]{Estimating Photometric Redshift from Mock Flux for CSST Survey by using Weighted Random Forest}

\author[J. Lu et al.]{
Junhao Lu$^{1}$,
Zhijian Luo$^{1}$\thanks{E-mail: zjluo@shnu.edu.cn},
Zhu Chen$^{1}$\thanks{E-mail: zhunchen@staff.shnu.edu.cn},
Liping Fu$^{1}$\thanks{E-mail: fuliping@shnu.edu.cn},
Wei Du$^{1}$,
Yan Gong$^{2,3}$,
Yicheng Li$^{1}$,
\newauthor
Xian-Min Meng$^{2}$,
Zhirui Tang$^{1}$,
Shaohua Zhang$^{1}$,
Chenggang Shu$^{1}$,
Xingchen Zhou$^{2,4}$
and Zuhui Fan$^{5,6}$
\\
$^{1}$Shanghai Key Lab for Astrophysics, Shanghai Normal University, Shanghai 200234, China\\
$^{2}$Key Laboratory of Space Astronomy and Technology, National Astronomical Observatories, Chinese Academy of Sciences,\\
20A Datun Road, Beĳing 100101, China\\
$^{3}$Science Center for China Space Station Telescope, National Astronomical Observatories, Chinese Academy of Sciences,\\ 
20A Datun Road, Beĳing 100101, China\\
$^{4}$University of Chinese Academy of Sciences, Beĳing 100049, China\\
$^{5}$South-Western Institute for Astronomy Research, Yunnan University, Kunming 650500, China\\
$^{6}$Department of Astronomy, School of Physics,Peking University, Beijing 100871, China\\
}

\date{Accepted 2023 December 19. Received 2023 December 19; in original form 2023 July 25}

\pubyear{2015}

\begin{document}
\label{firstpage}
\pagerange{\pageref{firstpage}--\pageref{lastpage}}
\maketitle

\begin{abstract}
Accurate estimation of photometric redshifts (photo-$z$) is crucial in studies of both galaxy evolution and cosmology using current and future large sky surveys. In this study, we employ Random Forest (RF), a machine learning algorithm, to estimate photo-$z$ and investigate the systematic uncertainties affecting the results. Using galaxy flux and color as input features, we construct a mapping between input features and redshift by using a training set of simulated data, generated from the Hubble Space Telescope Advanced Camera for Surveys (HST-ACS) and COSMOS catalogue, with the expected instrumental effects of the planned China Space Station Telescope (CSST). To improve the accuracy and confidence of predictions, we incorporate inverse variance weighting and perturb the catalog using input feature errors. Our results show that weighted RF can achieve a photo-$z$ accuracy of $\rm \sigma_{NMAD}=0.025$ and an outlier fraction of $\rm \eta=2.045\%$, significantly better than the values of $\rm \sigma_{NMAD}=0.043$ and $\rm \eta=6.45\%$ obtained by the widely used Easy and Accurate Zphot from Yale (EAZY) software which uses template-fitting method. Furthermore, we have calculated the importance of each input feature for different redshift ranges and found that the most important input features reflect the approximate position of the break features in galaxy spectra, demonstrating the algorithm's ability to extract physical information from data. Additionally, we have established confidence indices and error bars for each prediction value based on the shape of the redshift probability distribution function, suggesting that screening sources with high confidence can further reduce the outlier fraction.
\end{abstract}




\section{Introduction}
Photometric surveys are a powerful tool for investigating the properties of dark energy and dark matter, and for understanding the formation and evolution of cosmic large-scale structures (LSS) \citep{abdalla2011comparison,zhou2021spectroscopic}. Typically, late-time cosmological measurements are obtained by carefully measuring the three-dimensional distribution of galaxies using spectroscopic redshift (spec-$z$) \citep{carrasco2013tpz,cole2005power,percival2007shape}. However, spectroscopy that has high wavelength resolution requires more observation time to achieve sufficient signal-to-noise over a wide wavelength range \citep{salvato2019many}. Recently, more studies are relying on less precise statistical estimates of redshifts based on broadband photometry, known as photometric redshift (photo-$z$) \citep[e.g.][]{koo1985optical,loh1986photometric,wolf2001multi,arnouts2002measuring,collister2004ANNz,babbedge2004IMPZ,feldmann2006zurich,rowan2008photometric,hildebrandt2008blind,coupon2009photometric,soo2018morpho}. These estimation techniques have become crucial for multi-band digital surveys, allowing cosmological measurements on galaxy samples that are at least a hundred times larger than comparable spectroscopic samples \citep[e.g.][]{colless2001spectra,dickinson2003great,Skrutskie2006two,garilli2008vimos,garilli2014vipers,Wright2010wise,Grogin_2011_candles,Parkinson2012wigglez,chambers2016panstarrs,aihara2017hsc,zou2019photometric}. Additionally, photo-$z$ approaches have simple and uniform selection functions, extend to fainter flux limits and larger angular scales, and thus probe much larger cosmic volumes \citep{hildebrandt2010phat}.

There are numerous ongoing and next-generation photometric surveys underway or set to begin in the near future. These include the Sloan Digital Sky Survey (SDSS)\footnote{\url{http://www.sdss.org/}} \citep{fukugita1996sloan,york2000sloan}, Dark Energy Survey (DES)\footnote{\url{https://www.darkenergysurvey.org/}} \citep{collaboration2016more,abbott2021dark}, the Large Synoptic Survey Telescope (LSST)\footnote{\url{https://www.lsst.org/}} \citep{ivezic2019lsst,abell2009lsst}, the Euclid Space Telescope\footnote{\url{https://www.euclid-ec.org/}} \citep{laureijs2011euclid}, the Wide Field Infrared Survey Telescope (WFIRST)\footnote{\url{https://roman.gsfc.nasa.gov/}} \citep{spergel2015widefield}, and more. These surveys will detect billions of galaxies in a large redshift range through spectroscopic or photometric images, making it possible to accurately measure the dynamic evolution of the universe.

Currently, photo-$z$ computation methods can be broadly categorized into two categories: template fitting algorithms and empirical training algorithms, and many works have been done to compare these two different algorithms \citep[e.g.][]{hildebrandt2010phat,abdalla2011comparison,sanchez2014photometric,beck2017realistic,euclid2020photometric}. Template fitting algorithms \citep[e.g.][]{benitez2000bayesian,bolzonella2000photometric,csabai2003application,ilbert2006accurate,feldmann2006zurich,assef2010low} use either empirical \citep[e.g.][]{coleman1980colors,assef2010low} or synthetic spectral templates \citep[e.g.][]{bruzual2003stellar,maraston2005evolutionary,conroy2009propagation,eldridge2009spectral} to estimate photo-$z$. These techniques find the best match between the observed magnitudes or colors and the synthetic magnitudes or colors from the templates that are sampled across the expected redshift range of the photometric observations. Empirical training methods use a spectroscopic training dataset to calibrate an algorithm that can be quickly applied to new photometric observations. Initially, the training set was used to map a polynomial function between the colors or other photometric observables and the redshifts \citep[e.g.][]{connolly1995slicing,brunner1997toward,oyaizu2008galaxy,geach2012unsupervised,way2012can,sadeh2016ANNz2,tanaka2018photometric}. More recently, this process has been extended to machine learning algorithms such as Random Forest.

Previous works have utilized prediction trees in photo-$z$ calculation, such as \citet{carliles2010random}, who used the RF package in R to predict photo-$z$ and its error. They tested their approach on a subset of the SDSS Data Release 6 \citep{2008ApJS..175..297A} catalog with colors as input features and found that the RF method is very suitable for photo-$z$ prediction, producing comparable results to other machine learning methods. \citet{carrasco2013tpz} built upon these findings and developed a new parallel machine learning photo-$z$ Python code called TPZ\footnote{\url{http://matias-ck.com/mlz/index.html}}. TPZ employs both classification and regression trees in RF to calculate the probability density function (PDF). This approach utilizes extra information encoded within the measurement errors, generates ancillary information describing the spectroscopic training sample, and provides better control of the uncertainties. The authors tested their codes on galaxy samples drawn from the SDSS main galaxy sample and from the DEEP2 survey, obtaining excellent results in each case. \citet{fotopoulou2018cpz} combined three RFs and template fitting methods to identify stars, estimate redshifts for all galaxy populations including active galactic nucleis (AGN) and  quasi-stellar objects (QSO). They applied their codes on the near-infrared VISTA public surveys, matching them with optical photometry data from CFHTLS, KIDS, and SDSS, mid-infrared photometry from WISE, and ultraviolet photometry from the Galaxy Evolution Explorer (GALEX). Their analysis demonstrated that their methods enhance photometric redshift accuracy for both normal galaxies and AGN  without the need for extra X-ray information. \citet{zhou2019deep} assessed their cross-matched catalogue, combining $ugriz$ photometry from the Canada–France–Hawaii Telescope Legacy Survey (CFHTLS) and Y-band photometry from the Subaru Suprime camera, using a machine learning photometric redshift algorithm based upon RF regression. Their investigation revealed that utilizing corrected aperture photometry resulted in a notable enhancement in photo-$z$ accuracy when compared to the original SEXTRACTOR catalogues from CFHTLS and Subaru. \citet{mucesh2021machine} employed the DES catalog, incorporating redshift and stellar mass data from COSMOS2015, to assess the effectiveness of their python package ( GALPRO\footnote{\url{https://galpro.readthedocs.io/}}) based on RF. Their results revealed that their approach surpasses template fitting across all their predefined performance metrics. \citet{li2023desi} conducted a comparative analysis of three machine learning techniques, specifically CATBOOST, Multi-Layer Perceptron, and RF, in the context of cross-matched datasets involving the DESI Legacy Imaging Surveys DR9 galaxy catalog, as well as LAMOST DR7, GAMA DR3, and WiggleZ galaxy catalogs. Although their findings indicate that RF is not the most optimal method for redshift estimation in their study, it exhibited a more favorable bias compared to CATBOOST.

Given its simplicity and robust performance in photometric redshift estimation, RF stands out as a valuable tool for showcasing feature importance, offering valuable insights for future survey strategies. However, the original RF algorithm proved to be ineffective when applied directly. As a response, enhancements were made to the algorithm, leading to a more effective and improved methodology for achieving better results. It is worth noting that there are also other powerful algorithms available, such as CATBOOST. In this work, we utilize RF to estimate the accuracy of photo-$z$ in the optical survey of CSST, a 2-meter diameter space telescope expected to launch around 2024 and will be in the same orbit as the China Manned Space Station \citep{zhan2011consideration,cao2018testing,gong2019cosmology}.

\begin{figure}
	\includegraphics[width=\columnwidth]{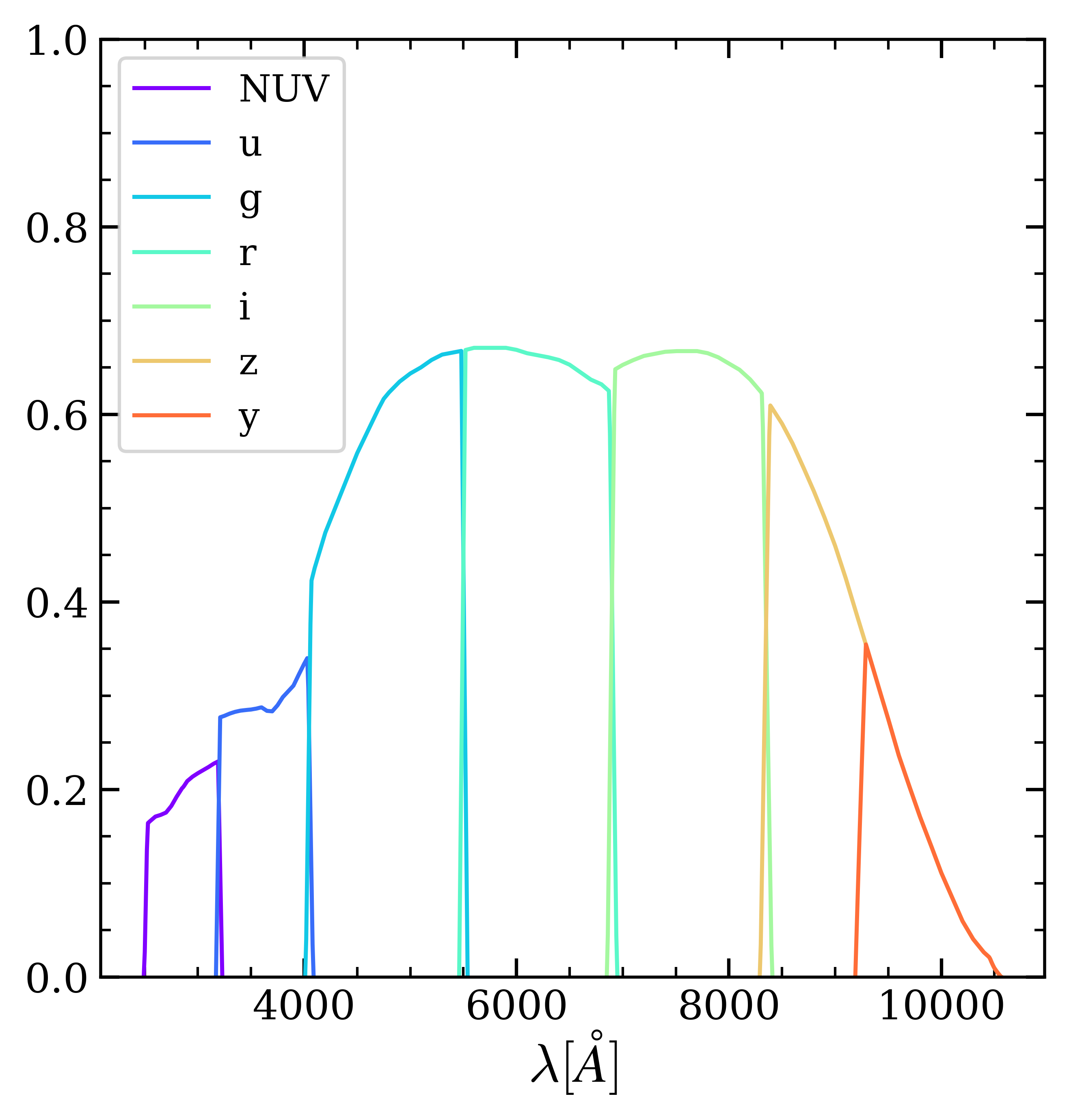}
	\caption{The solid curves represent the real transmissions of the CSST seven photometric bands. The effect of detector quantum efficiency is included. The details of the transmission parameters can be found in \citet{cao2018testing}}
	\label{fig:filters}
\end{figure}

CSST is planned to observe about 17,500 deg$^2$ in approximately 10 years, covering the optical and NIR bands from $\sim250$ nm to $\sim1000$ nm. The $5\sigma$ limit for a point source magnitude can be as high as $\sim26$ AB mag for the $g$, $r$, and $i$ bands, and is approximately $24.5\sim25.5$ for the other bands. In Figure \ref{fig:filters}, the real transmissions, including detector quantum efficiency, of the CSST seven photometric filters are shown. The primary scientific goals of CSST involve exploring the evolution of LSS, the properties of dark matter and dark energy, galaxy formation and evolution, among others, hence, photo-$z$ measurements are necessary.

In a recent study, \citet{zhou2022photometricBNN} explored the accuracy of photo-$z$ methods in the CSST using four neural networks: Multilayer Perceptron (MLP), Convolutional Neural Network (CNN), Hybrid, and Hybrid transfer. These networks were used to extract photo-$z$ information from mock CSST flux and image data, achieving a photo-$z$ accuracy of approximately 0.02 and outlier fraction of around 0.9\%.

As part of a series of studies on CSST photo-$z$ estimation, we tested the use of the RF algorithm. We generated mock photometric training data from the COSMOS galaxy catalog \citep{capak2007first,ilbert2008cosmos} incorporating CSST instrument effects. Our aim is to assess the accuracy of this method and evaluate the credibility of prediction results. We also attempted to improve the prediction process by adding errors of input feature as fitting weight. These tests will inform the development of strategies for estimating photo-$z$ using RF on real CSST data in the future.

The paper is structured as follows: Section \ref{section:mock} describes the generation of mock flux for training, while Section \ref{sec:methods} introduces the EAZY and RF algorithm for the redshift estimation problem. In Section \ref{section:results}, we apply RF algorithm and EAZY code to the mock data and present the results. Finally, we summarize the study in Section \ref{section:conclusion}.

\section{mock data}\label{section:mock}

In this section, we will briefly introduce how the mock data is generated. Further details about the mock process can be found in \citet{zhou2021spectroscopic}. The mock data should have similar properties like redshift, magnitude distribution and galaxy type etc. with observations of the CSST survey. To achieve a high degree of realism in simulating galaxy images for the CSST photometric survey, we employ mock image generation techniques rooted in observations taken within the COSMOS field using the Advanced Camera for Surveys of the Hubble Space Telescope (HST-ACS), incorporating CSST instrumental effects. The mock flux data of galaxies are measured from these images by aperture photometry. The HST-ACS survey covers an area of approximately 2 deg$^2$ in the F814W band, which has a spatial resolution similar to that of the CSST, with an 80\% energy concentration radius of $R_{80}\sim0.^{\prime\prime}15$ \citep{cao2018testing,gong2019cosmology,koekemoer2007cosmos,massey2010pixel,bohlin2016perfecting}. Also the COSMOS HST-ACS F814W survey boasts a notably low background noise, anticipated to be approximately 1/3 of that in the CSST survey. This characteristic makes it a solid basis for simulating CSST galaxy images.

We first select the central $0.85\times0.85$ deg$^2$ area of this survey containing $\sim$ 192,000 galaxies to obtain high-quality images. Then rescale the pixel size of the COSMOS HST-ACS F814W survey from $0.03''$ to $0.075''$ to match the CSST pixel size. Next we extract a square stamp image for each galaxy from the survey area, with a galaxy in the center of the stamp image. The square stamp's dimensions are 15 times that of the galaxy's semi-major axis, resulting in varying sizes for galaxy stamp images. The galaxy's semi-major axis and other pertinent morphological details can be found in the COSMOS weak lensing source catalog \citep{Leauthaud2007weak}. In addition, we mask all sources in the stamp image with a signal-to-noise ratio (SNR) greater than $3\sigma$, except for the central galaxy, and substitute them with the CSST background noise.

Here we use the COSMOS2015 catalog \citep{COSMOS2015} to match galaxies in HST-ACS F814W survey, it contains about 220,000 galaxies with the measures of galaxy redshift, magnitude, size, dust extinction, best-fit spectral energy distribution (SED), and so on \citep{capak2007first,ilbert2008cosmos,cao2018testing,gong2019cosmology}. The COSMOS photo-$z$s have been computed using more than 30 bands that span a wide range of the electromagnetic spectrum. \citet{COSMOS2015} have conducted verification of some of the photo-$z$ estimates in the COSMOS2015 catalog by comparing them with several spectroscopic survey samples. The accuracy of photo-$z$s and characteristics of the spec-$z$ samples can be found in Tables 4 and 5, as well as Figures 11 and 12 presented by \citet{COSMOS2015}. Given that the photo-$z$s in this catalog have demonstrated precision and accuracy, we consider these photo-$z$ estimates to be reliable and have chosen to adopt them as the true redshift (hereafter referred to as $z_{\rm true}$) for the purpose of training RF. However, it should be noted that due to this reason, there may be some potential bias or error in the present results. In the future, the same method can be applied with the spectroscopic sample obtained in the CSST survey. In this work, 31 SED templates from $LePhare$ code \citep{arnouts1999measuring,ilbert2006accurate} were reproduced to fit galaxy flux. We extend the wavelength range of the templates from $\sim900\textup{\AA}$ to $\sim90\textup{\AA}$ using the BC03 method \citep{bruzual2003stellar} because CSST has large wavelength coverage between NUV and NIR band. During the process of fitting the flux with the SED templates, we include five dust extinction laws, derived from sources like the Milky Way \citep{allen1976astrophysical,Seaton1979interstellar}, the Large Magellanic Cloud \citep{Fitzpatrick1986average}, the Small Magellanic Cloud \citep{Prevot1984typical,bouchet1985visible}, and a starburst galaxy \citep{Calzetti2000dust}, as well as various emission lines such as Ly$\alpha$, H$\alpha$, H$\beta$, [O\uppercase\expandafter{\romannumeral2}], and [O\uppercase\expandafter{\romannumeral3}]. For the IGM absorption, we use the attenuation laws computed by \citet{madau1995radiative}.

Then we convolve the obtained SEDs with the CSST total transmission curves (Figure \ref{fig:filters}) to calculate the theoretical flux data observed by the CSST. The theoretical flux in electron counting rate of a band $j$ can be estimated as:
\begin{equation}\label{eq:eq1}
	C_{\rm th}^j=A_{\rm eff}^j\int S(\lambda)\tau_j(\lambda)\frac{\lambda}{hc}d\lambda,
\end{equation}
where $h$ and $c$ are the Planck constant and speed of light respectively, $A$ is the effective telescope aperture area, $S(\lambda)$ is the  SED, and $\tau_j(\lambda)=T_j(\lambda)Q_j(\lambda)M_j(\lambda)$ is total system throughput, here $T_j(\lambda)$,$ Q_j(\lambda)$ and $M_j(\lambda)$ are the intrinsic filter transmission, detector quantum efficiency, and total mirror efficiency respectively. Then we can calculate the theoretical flux in electron counts of $j$ band as:
\begin{equation}
	F_{\rm th}^j=C_{\rm th}^jt_{\rm exp}N_{\rm exp}^j,
\end{equation}
where $t_{\rm exp}=150s$ is exposure time, $N_{\rm exp}^j$ is the number of exposures, for $NUV$ and $y$ bands, it is 4, for the other bands it is 2.

After that, the CSST instrumental effects and background noise were added to the image. The additional noise in $j$ band can be expressed as
\begin{equation}\label{eq:noise}
	N_{\rm add}^j=\sqrt{(N_{\rm bkg,th}^j)^2-(N_{\rm img}^j)^2},
\end{equation}
where $N_{\rm img}^j$ is the background noise of the rescaled image, $N_{\rm bkg,th}^j$ is the theoretical CSST background noise per pixel, it can be calculated by
\begin{equation}
 N_{\rm bkg,th}^j=\sqrt{(B_{\rm sky}^j+B_{\rm dark})t_{\rm exp}N_{\rm exp}^j+R_n^2N_{\rm exp}^j},
\end{equation}
here $B_{\rm dark}=0.02\rm e^-s^{-1}pix^{-1}$ is the dark current,$ R_n= 5\rm e^-pix^{-1} $is the read-out noise, and $B_{\rm sky}^j$ is the sky background in unit of $\rm e^-s^{-1}pix^{-1}$ ,which is given by
\begin{equation}
	B_{\rm sky}^j=A_{\rm eff}^jl_{\rm pix}^2\int I_{\rm sky}(\lambda)\tau_j(\lambda)\frac{\lambda}{hc}d\lambda,
\end{equation}
where $I_{\rm sky}(\lambda)$ is the surface brightness intensity of the sky background in units of $\rm erg\cdot s^{-1}cm^{-2}\textup{\AA}^{-1}arcsec^{-2}$, and $l_{\rm pix} $is the pixel size in arcseconds. The evaluation of $B_{\rm sky}^j$ is based on the measurements of the earthshine and zodiacal light for the ‘average’ sky background case given in \citet{ubeda2011acs}. It is found that $B_{\rm sky}^j$ are 0.0023, 0.0018, 0.142, 0.187, 0.187, 0.118 and 0.035 $\rm e^-s^{-1}pix^{-1}$ for $NUV$, $u$, $g$, $r$, $i$, $z$ and $y$ bands. Then $N_{\rm bkg,th}^j$ are calculated as 10.65, 7.84, 9.93, 10.59, 10.59, 9.56 and 11.53 $\rm e^-$ for these bands. Thus the additional noise $N_{\rm add}^j$ can be calculated by subtracting the rescaled images noise $N_{\rm img}^j$ as shown in equation \ref{eq:noise}, and added to pixels in stamp images by sampling from a Gaussian distribution with $\rm mean=0$ and $\sigma=N_{\rm add}^j$ in band $j$.

After obtaining the final mock image, aperture photometry method was used to measure the flux. First, $g$, $r$, $i$ and $z$ band were staked to create detection images for high SNR source. Then we measure the Kron radius \citep{kron1980photometry} along galaxy major- and minor-axis to find an elliptical aperture with the size $1\times R_{\rm Kron}$. Finally, we can get the flux and error from electron number $N_{\rm e^-}^j$ and error $\sigma_{\rm e^-}^j$ measured within the aperture. The SNR in band $j$ can be calculated as
\begin{equation}
	{\rm SNR}_j=\frac{F_{\rm obs}^j}{\sqrt{F_{\rm obs}^j+N_{\rm pix}(B_{\rm sky}+B_{\rm dark})t_{\rm exp}N_{\rm exp}^j+R_n^2N_{\rm exp}^j N_{\rm pix}}},
\end{equation}
where $F_{\rm obs}^j$ is the observed electron counts, and $N_{\rm pix}$ is the number of pixels covered by a galaxy for the CSST. In this way, we have constructed a mock data sample for CSST photometric observations.

\begin{figure}
	\includegraphics[width=\columnwidth]{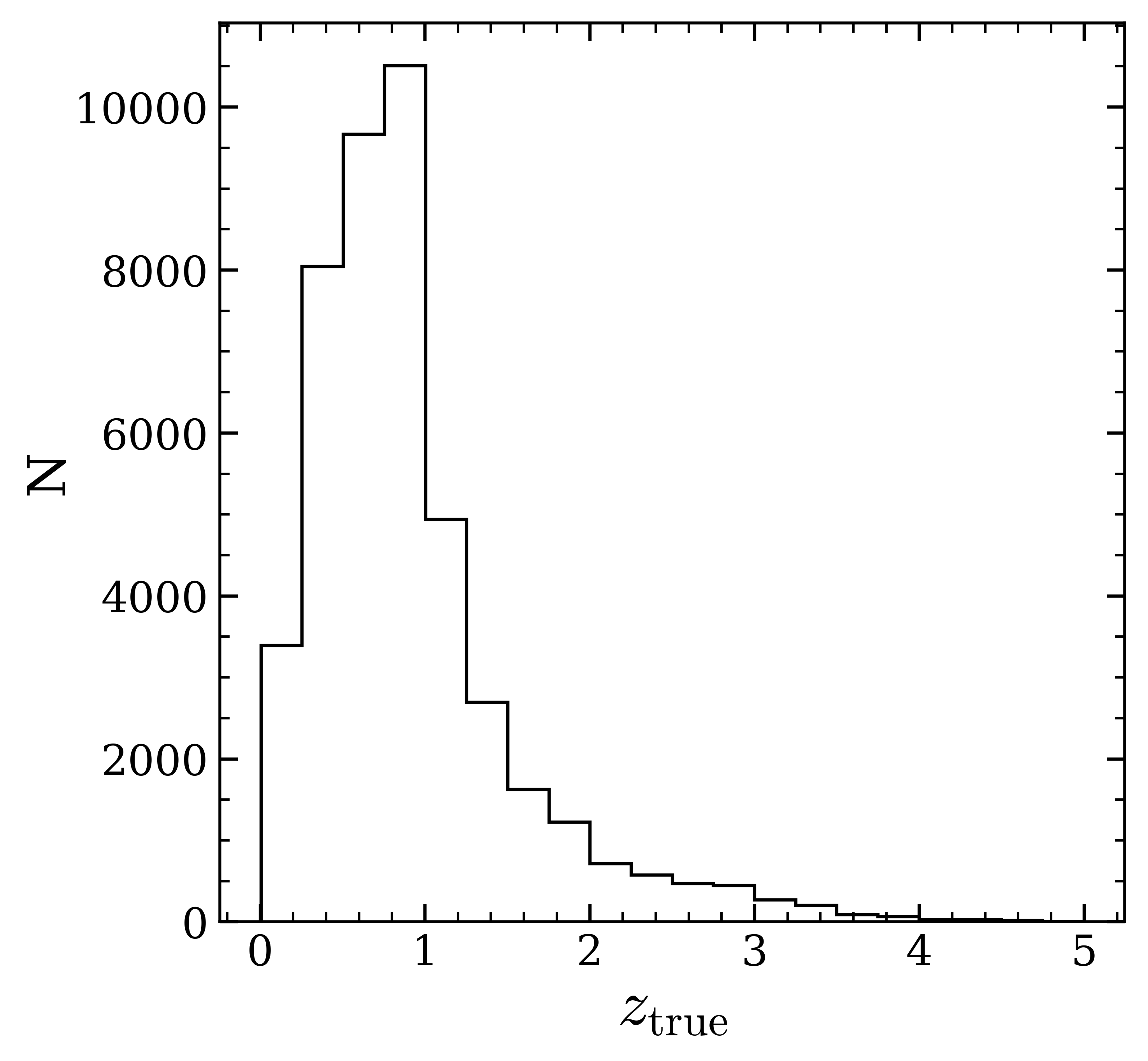}
	\caption{Galaxy redshift distribution of the mock sample from the COSMOS catalog. The sources have been selected with SNR greater than 10 in g or i bands. The distribution ranges from 0 to $\sim5$ with a peak locates around $0.6-1$.}
	\label{fig:redshift}
\end{figure}

\begin{figure}
	\includegraphics[width=\columnwidth]{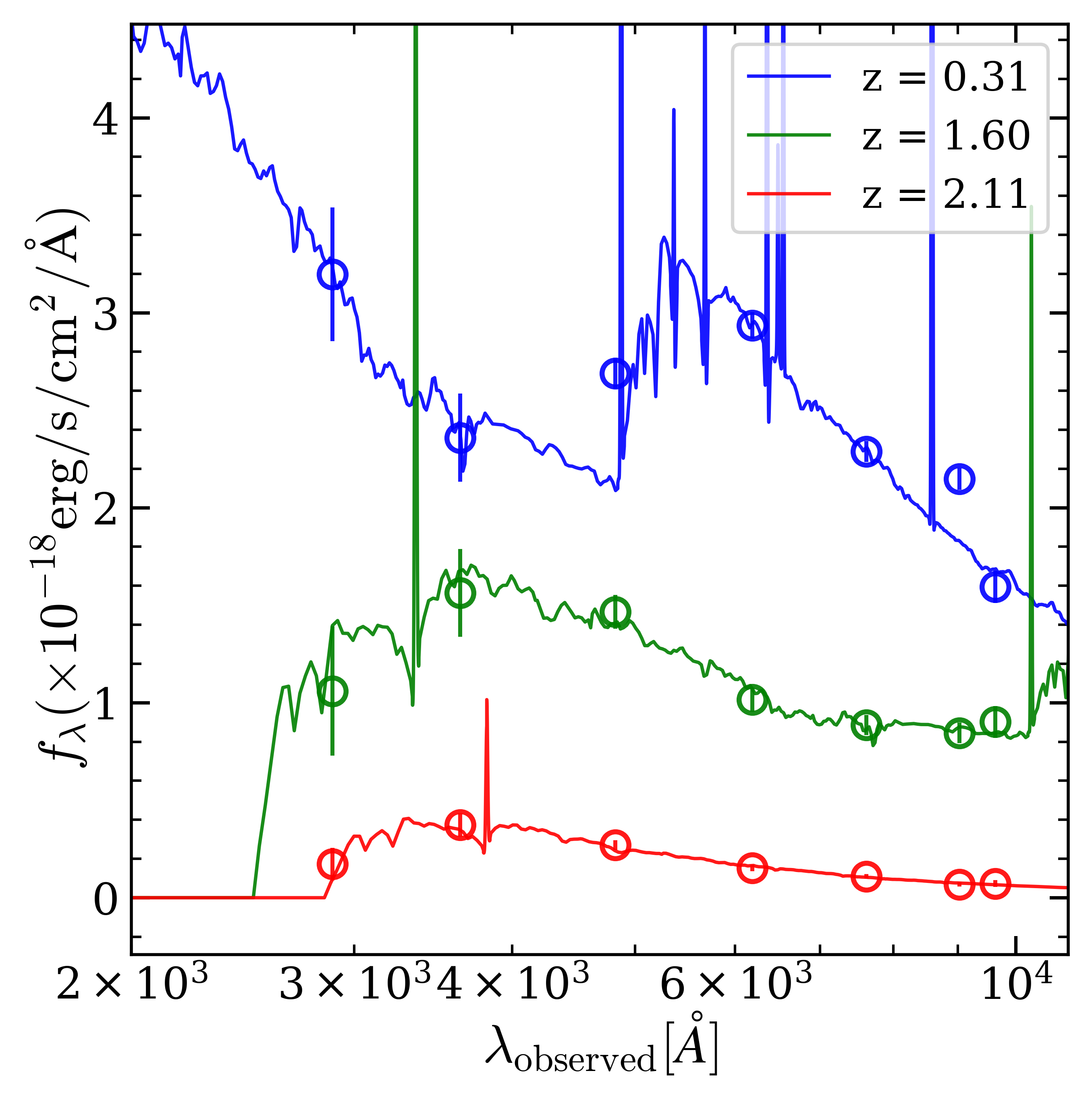}
	\caption{The fluxes of galaxy samples measured by the aperture photometry method in seven bands. The SEDs of the corresponding galaxy are also shown in solid curves as comparison.}
	\label{fig:sed}
\end{figure}

Figure \ref{fig:redshift} displays the redshift distribution of the mock data. The peak of the distribution is around $z=0.6-1$, covering the redshift range from 0 to 5.
Figure \ref{fig:sed} presents measured flux examples from mock images for the seven CSST photometric bands at $z=0.31$, $1.6$ and $2.52$, accompanied by the corresponding SEDs for comparison. The figure clearly shows that the mock flux data accurately captures the features of galaxy SEDs.

\section{Methods}\label{sec:methods}

\subsection{EAZY approach}
EAZY, as described in \citet{brammer2008eazy}, is a photometric redshift software that employs a template-fitting approach. The fundamental procedure involves the minimization of the $\chi^2$ statistic by comparing SED templates with observed colors across various redshift grids:
\begin{equation}
	\chi_{z,i}^2=\sum_{j=1}^{N_{\rm filt}}\frac{\left(T_{z,i,j}-F_j\right)^2}{\left(\delta F_j\right)^2},
	\label{eq:chisquare}
\end{equation}
where $N_{\rm filt}$ is the number of filters, $T_{z,i,j}$ is the synthetic flux of template $i$ in filter $j$ for redshift $z$, $F_j$ is the observed flux in filter $j$, and $\delta F_j$ is the uncertainty in $F_j$. Then EAZY will find the best-fitting coefficients $\alpha_i$ of all combined templates:
\begin{equation}
	T_z=\sum_{i=1}^{N_{\rm temp}}\alpha_iT_{z,i}.
\end{equation}

Among all template-fitting methods, EAZY is one of the most widely used software. For example, \citet{yang2014photometric} employed EAZY to derive photo-$z$s for the Hawaii-Hubble Deep Field-North (H-HDF-N) survey catalog, and \citet{chen2018xmm} estimated photo-$z$s for the X-ray point-source catalogue within the XMM-Large Scale Structure (XMM-LSS) survey region using EAZY. Both studies demonstrated strong performance of EAZY. Additionally, \citet{euclid2020photometric} showed that when run in identical configuration, template-fitting methods provide nearly identical results. The differences observed in the results are not due to differences in performance of the template-fitting methods, but rather to variations in their configurations. Hence, we have selected the EAZY method as a representative template-fitting approach for conducting a comparative analysis with our RF algorithm.

For EAZY, we employed the default EAZY\_v1.2\_dusty templates, which include 8 templates and were widely used. Additionally, we incorporated the $r$ band apparent magnitude prior $p(z|m_r)$, which represents the redshift distribution of galaxies with the apparent magnitude $m_r$. The parameter $Z\_STEP\_TYPE$ is configured with a value of 0, indicating an even division of the redshift grid. All other parameters are retained at their default settings. The input bands consist of the seven CSST bands described before.

\subsection{Random forest algorithm}\label{section:rf}
Random Forest \citep{breiman2001random} is an ensemble learning algorithm that consists of multiple regression trees trained on bootstrap samples \citep{caruana2008empirical}. In this work, we utilize the RF class from the scikit-learn library \footnote{\url{https://scikit-learn.org/stable/}}\citep{scikit-learn} to import and apply the Random Forest algorithm. Our goal is to train independent prediction trees on the mock data to accurately estimate photo-$z$, and to use the resulting probability density function to characterize the error and derive an estimation for each galaxy. We also calculate other useful information such as feature importance and redshift confidence following the method proposed in \citet{carrasco2013tpz}. The details of this algorithm and training process can be seen in Appendix \ref{app:rf}. 

\subsubsection{Practical details of the procedure}
In future analyses of weak gravitational lensing surveys, CSST will utilize photo-$z$ data. To ensure the accuracy of the photo-$z$ information, we have selected a high-quality sample from mock data with a signal-to-noise ratio (SNR) larger than 10 in either the $g$ or $i$ band. The resulting subsample contains 44,991 sources. In future work, other potentially better selection criteria such as size or shape measurement quality will be taken into consideration.

\begin{figure}
	\includegraphics[width=\columnwidth]{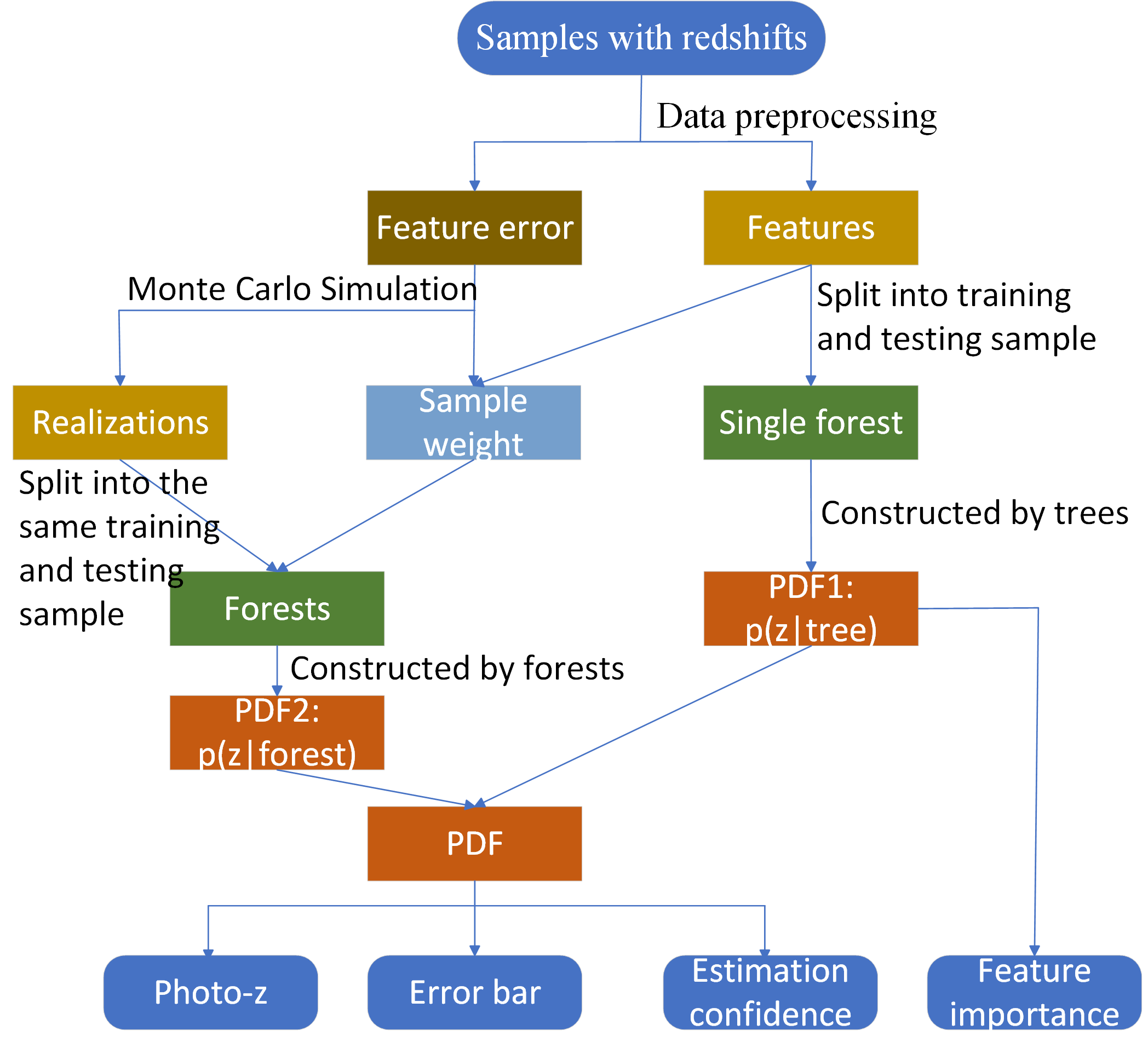}
	\caption{A simplified workflow of the whole algorithm. After preprocessing the sample with true redshifts, we can get the input features and their errors. Then using the Monte Carlo Simulation to expand the catalog. By combining the PDF from the trees in original catalog and from the forests in the mimic catalog, we obtain the last PDF. Our final photo-$z$ and error bar will be calculated based on it.}
	\label{fig:workflow}
\end{figure}

Figure \ref{fig:workflow} depicts a simplified workflow of the algorithm. First of all, the sample is randomly divided into training and testing data with a temporary ratio of 1:1. The input features consist of the flux of seven bands and six "colors" derived from the ratio of $\frac{{\rm flux}_i}{{\rm flux}_j}$, where ${\rm flux}_i$ and ${\rm flux}_j$ are adjacent. Our mock data does not contain missing values, but may occasionally be negative, as this can occur when sources are undetected in certain bands. 
These features are rescaled to the same level using standard scaler for normalization. The formula for standard scaler calculation is:
\begin{equation}
	x^*=\frac{x-\mu}{\sigma},
\end{equation}
where $\mu$ is the mean value of the data, $x$ is the value of every feature and $\rm \sigma$ is the standard deviation of all data. 

\begin{figure}
	\includegraphics[width=\columnwidth]{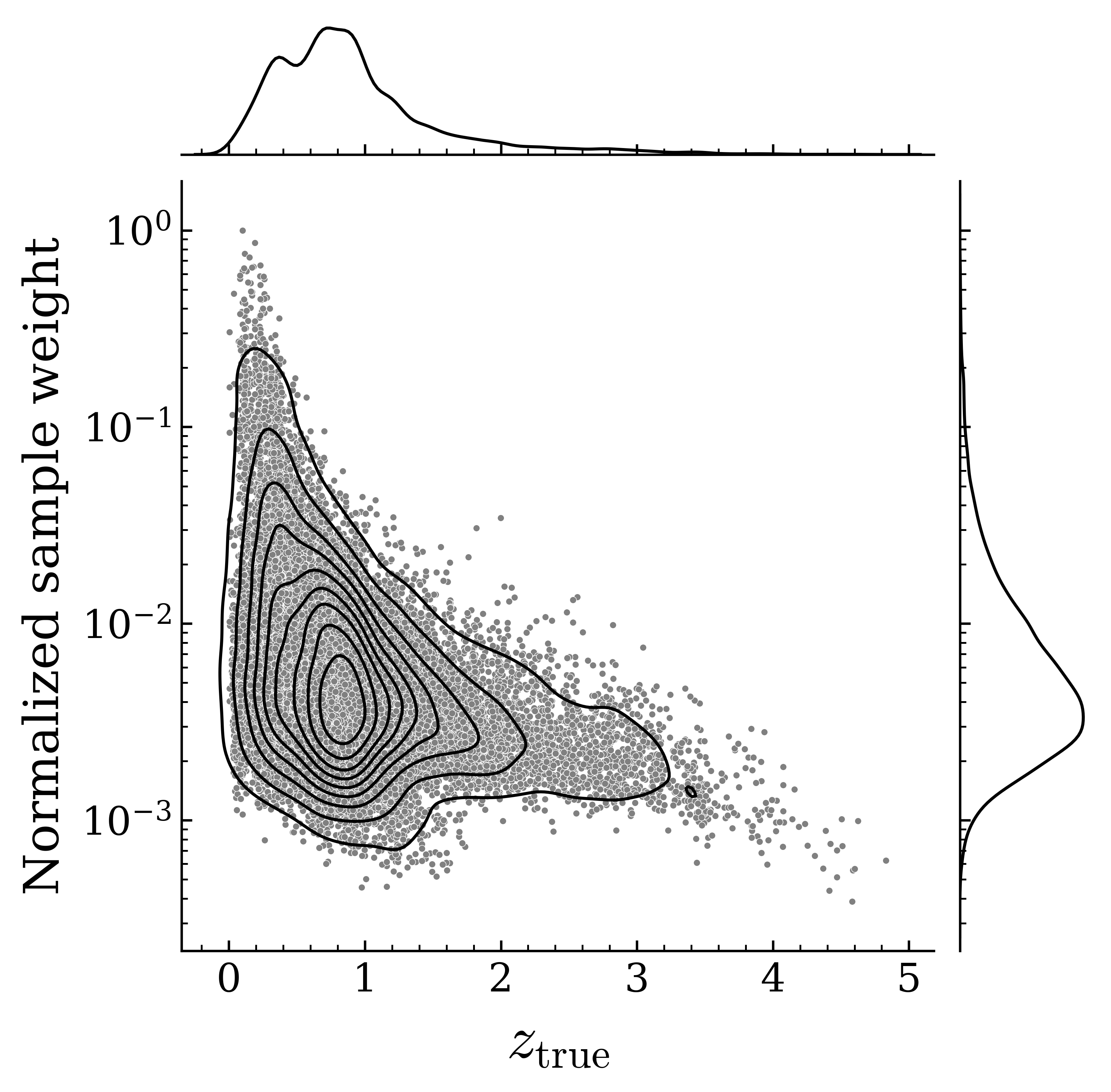}
	\caption{The normalized sample weight of each source (gray points) in different redshifts.  Most of the sources share similar weights. The density contour is shown with black lines. The distribution of redshifts and sample weights are plotted on the top and right panel, respectively.}
	\label{fig:sampleweight}
\end{figure}

Then, we have developed a method based on the RF algorithm to fully utilize the observation error information. Instead of using the observation error directly as input features, we first compute the inverse variance as the sample weight during the fitting process,
\begin{equation}
	 {\rm sample~weight} = \frac{\sum_jy_j/\sigma_j^2}{\sum_j1/\sigma_j^2},
\end{equation}
where $\sigma_j$ is the error of input features in $j$ band, $y_j$ is the corresponding feature. The sample weight for each source is shown in Figure \ref{fig:sampleweight}. From the figure, we observe that the weights are concentrated and decrease as redshift increases. 

The RF algorithm generates prediction results for all trees, which can be used to construct the Probability Density Function (PDF) of redshift through statistical analysis. \citet{carrasco2014sparse} and \citet{rau2015accurate} have explored various methods for generating PDFs with RF, such as the Gaussian mixture model (GMM) and non-parametric techniques like kernel density estimation (KDE) \citep{murray1956remarks,parzen1962estimation}, among others. Due to its simplicity, we opt for KDE to construct our PDFs. KDE operates by placing a kernel, typically a smooth, bell-shaped function, at each data point and summing these kernels to produce a smooth, continuous approximation of the probability density. As a non-parametric method, it estimates $\hat{p}(x)$ for the PDF $p(x)$ using a set of $N$ samples $x_i$. In the case of 1D KDE, the form is as follows:
\begin{equation}
	\hat{p}(x)=\frac{1}{nh}\sum_{i=1}^{n}K\left(\frac{x-x_i}{h}\right),
\end{equation}
where we define $K$ as the Gaussian kernel, with its standard deviation determined by the standard deviation of the $n$ samples, where $h$ is referred to as the bandwidth. For this purpose, we employ the $scipy.stats.gaussian\_kde$\footnote{\url{https://scipy.org/}} class. We retain all parameters at their default settings. It is worth noting that our PDF is not a conventional PDF, rather, it represents the distribution of predictions. This PDF is denoted as $p(z|{\rm tree})$, which represents the probability distribution of redshift. Typically, a redshift prediction can be either the mean value or the mode of the entire PDF. In this study, we use both for comparison. Our findings indicate that by selecting the peak redshift, the outlier fraction in test data can be reduced from about 4\% to 2.3\%. Therefore, we choose the peak value as the final prediction. In addition, the input feature importance can be obtained after constructing the RF. 

Next, we utilize a Monte Carlo simulation while taking the input feature error as a 1$\sigma$ standard deviation to expand the training set. This process generates a factor of 100 new mimic catalogs. Utilizing the above RF algorithm, we make redshift predictions for each catalog. These predictions are used to create a new PDF, called $ p(z|{\rm forest})$, which takes into account observation error.

After that, we construct a final PDF, which is obtained by $p(z)\propto p(z|{\rm tree})p(z|{\rm forest})$, to predict the redshift by selecting the value where the peak is located. Here the term $p(z|{\rm forest})$ can be regarded as a straightforward weighting system based on the model.

In addition, we define the outlier source according to
\begin{equation}
	\frac{|z_{\rm true}-z_{\rm phot}|}{1+z_{\rm true}}>0.15,
\end{equation}
then we use the outlier fraction and normalized median absolute deviation (NMAD) \citep{brammer2008eazy} 
\begin{equation}
	{\rm \sigma_{NMAD}}=1.48\times {\rm median}\left(\left|\frac{\Delta z - {\rm median}(\Delta z)}{1+z_{\rm true}}\right|\right),
\end{equation}
to judge the accuracy, where 
\begin{equation}
	\Delta z =z_{\rm phot}-z_{\rm true},
\end{equation}
$z_{\rm true}$ and $z_{\rm phot}$ are the real and predicted redshift respectively. This step mainly estimates the performance of the algorithm in redshift prediction.

Our final step is to establish additional indicators, which refer to \citet{carrasco2013tpz}, to evaluate the performance of our model predictions. Additionally, we investigate the effect of changing the ratio of training and testing data to determine the optimal amount of training data.

\begin{figure*}
	\centering
	\subfloat[]{\includegraphics[width=0.33\textwidth]{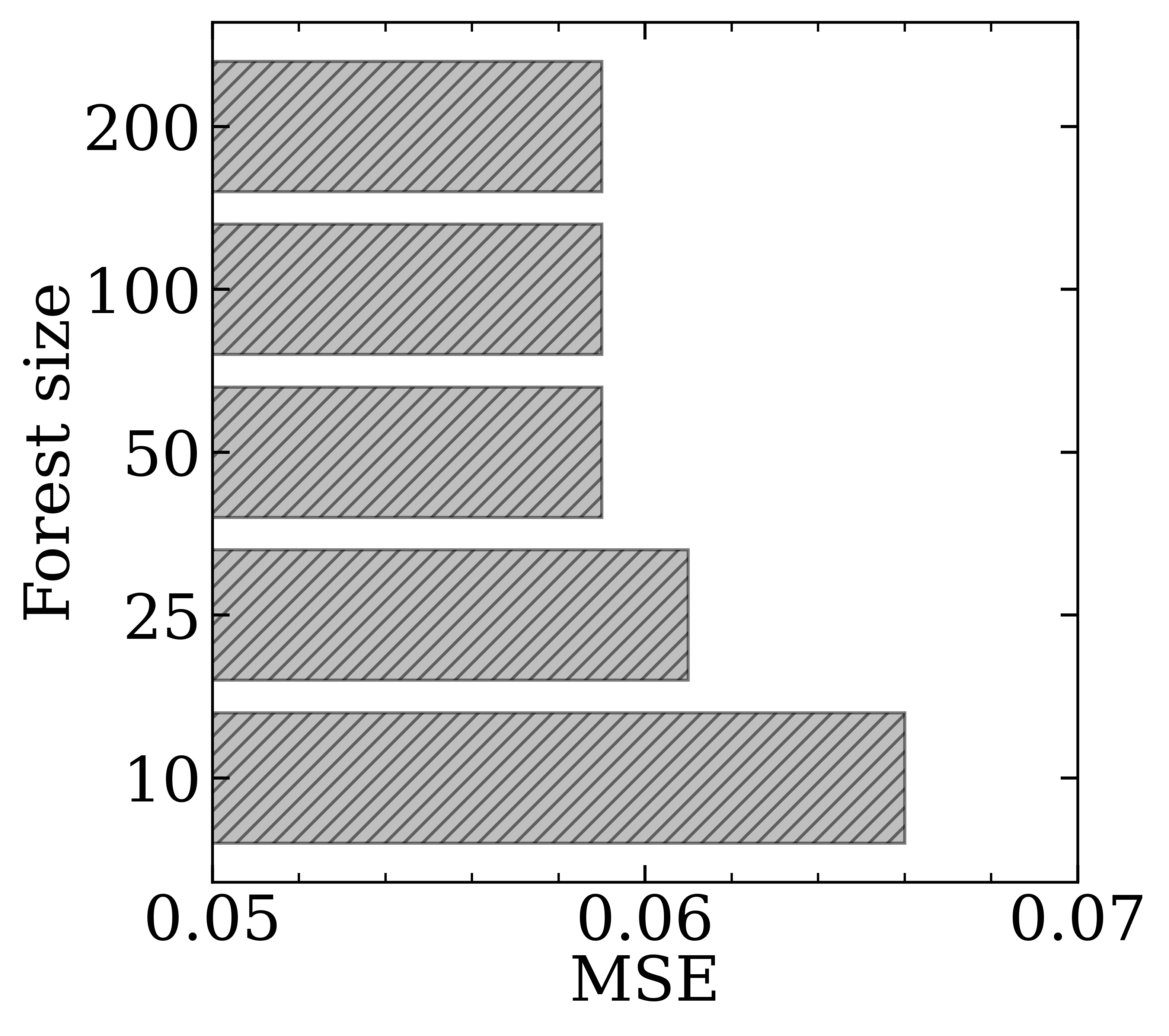}}
	\hfill
	\subfloat[]{\includegraphics[width=0.32\textwidth]{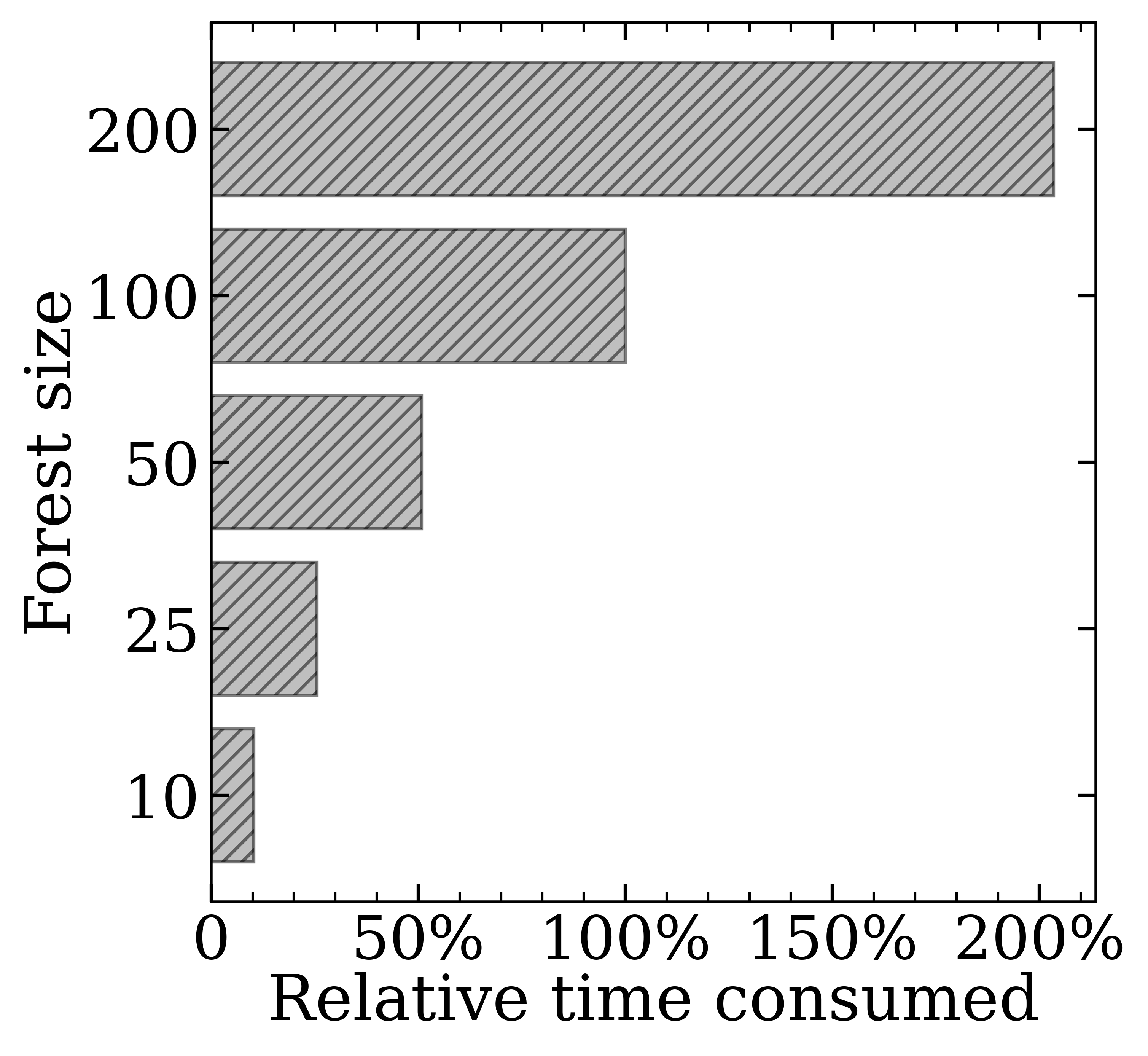}}
	\hfill
	\subfloat[]{\includegraphics[width=0.305\textwidth]{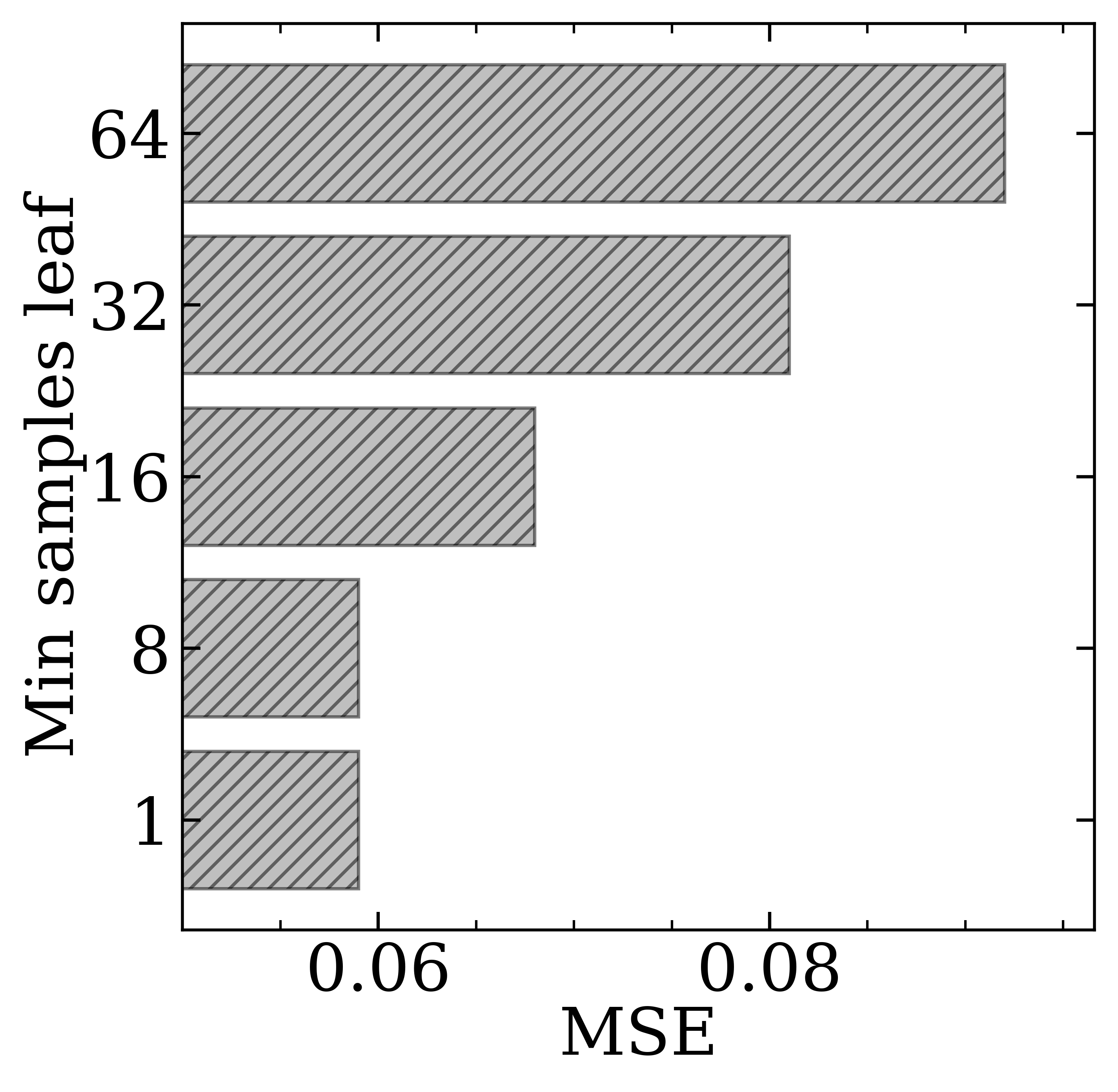}}
	\hfill
	\caption{(a).Mean squared error (MSE) with the forest size of 10, 25, 50, 100, 200. (b).Relative time consumed with different forest size. The time consumed with 100 trees is set as standard time ($100\%$). (c).MSE with min samples leaf of 1,8,16,32,64 when forest size is 100.}
	\label{fig:adjpara}
\end{figure*}

We also conducted separate tests to measure the impact of forest size and other parameters in the fitting process on performance. The number of regression trees in the forest is an essential hyperparameter that can affect the algorithm's performance. Generally, increasing the forest's size can improve the accuracy of the model as it increases the diversity of the trees and reduces the risk of overfitting. However, there is a point of diminishing returns beyond which further increasing the forest's size can not significantly improve performance but increase computational cost. In Figure \ref{fig:adjpara}(a), we demonstrate how the mean squared error (MSE) of the result is influenced by the forest's size. The MSE on the training data improves significantly when the forest size is small, but after 50, it changes little until it stops decreasing. On the other hand, the time required grows proportionally as the forest size increases. As shown in Figure \ref{fig:adjpara}(b), if we consider the time consumed in the model with 100 trees as the standard duration, models with 200 trees will cost twice as much time while the MSE changes little. Therefore, it is not necessary to establish too many trees.

Figure \ref{fig:adjpara}(c) illustrates the role of another parameter, the minimum number of samples required to be a leaf node, known as "min\_samples\_leaf". It determines the minimum number of samples that must be present in a leaf node of a regression tree before the splitting process can be halted. Setting a higher value for min samples leaf can lead to a simpler tree with fewer nodes, reducing the risk of overfitting the training data. It can also make the regression tree more interpretable by reducing the number of splits and resulting in a more intuitive regression process. Such modification is called "prune". However, if we cut too many nodes, the model will have the risk of underfitting. The result in Figure \ref{fig:adjpara}(c) is based on the forest size of 100. Other parameters, such as "max\_depth" and "min\_impurity\_decrease", have a similar role. By adjusting these parameters, it is possible to control the size, complexity, and generalization ability of the regression trees in the RF. In general, the optimal values for these parameters will depend on the specific problem being addressed and the desired trade-off between accuracy and model complexity. In this work, we find that setting them to their default values is good enough.

\section{Results}\label{section:results}
\begin{figure}
	\centering
	\includegraphics[width=\columnwidth]{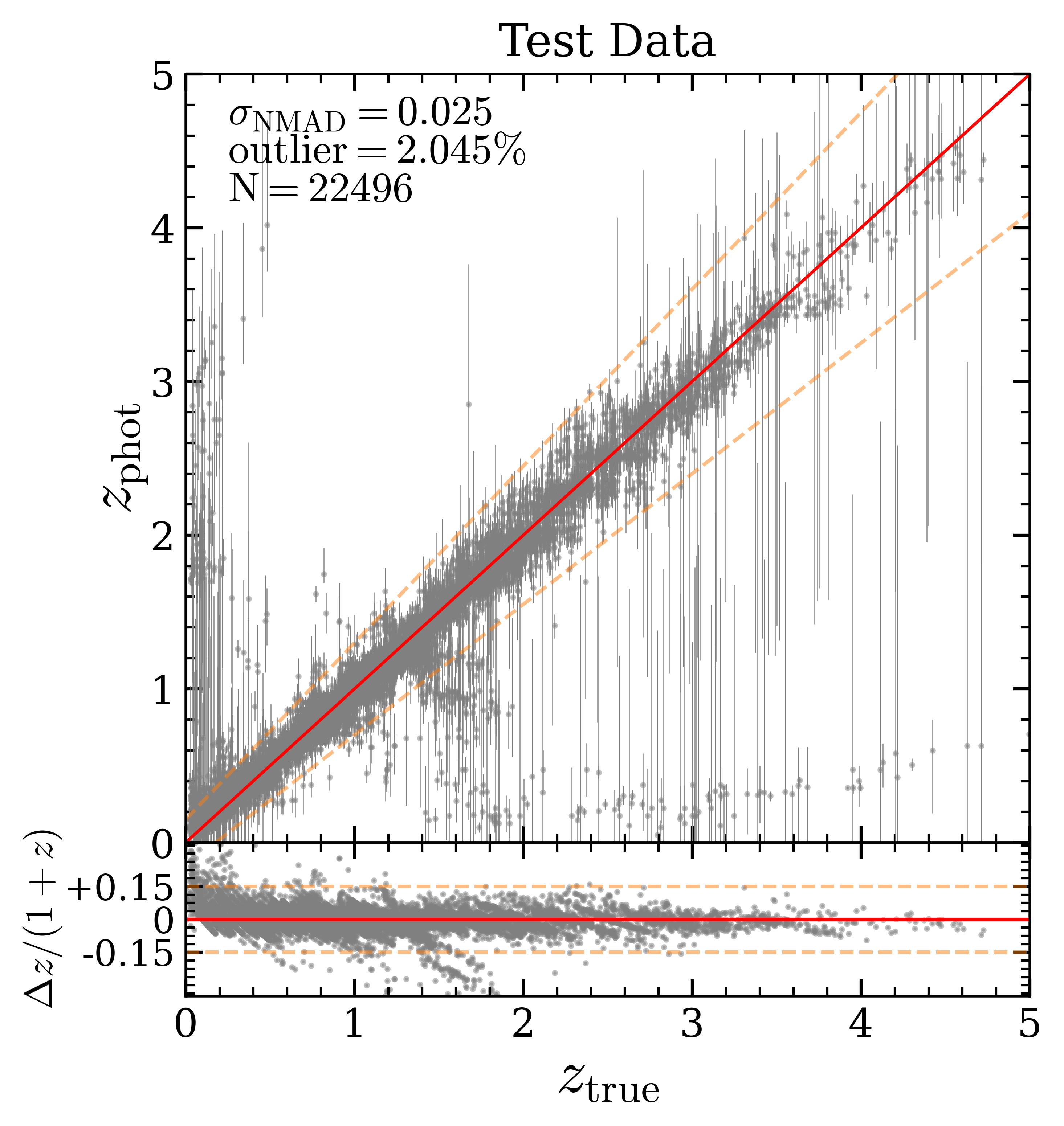}
	\caption{Error bar of using RF in testing data. The 16th and 84th percentile of the cumulative PDF are calculated as the upper and lower limits of the error bars. Point is the prediction value, also one of the peak value located in PDF.}
	\label{fig:confandbar}
\end{figure}

The results in Figure \ref{fig:confandbar} demonstrate that the outlier fraction and $\rm\sigma_{NMAD}$ in the testing data are 2.045\% and 0.025, respectively, which are similar to the results of 1.823\% and 0.021 in the training data. This indicates that the model does not suffer from overfitting. Moreover, our method outperforms the EAZY with a specific configuration, which yields an outlier fraction of 6.45\% and $\rm\sigma_{NMAD}$ of 0.043. Although some studies, such as \citet{euclid2020photometric}, suggest that certain model fitting methods are superior to the RF method, for our CSST photometric simulated sample, both in terms of the outlier fraction and $\rm \sigma_{NMAD}$, the RF method outperforms EAZY. The comparison between the RF method and other model fitting methods will be conducted in further research.

\begin{figure}
	\centering
	\includegraphics[width=\columnwidth]{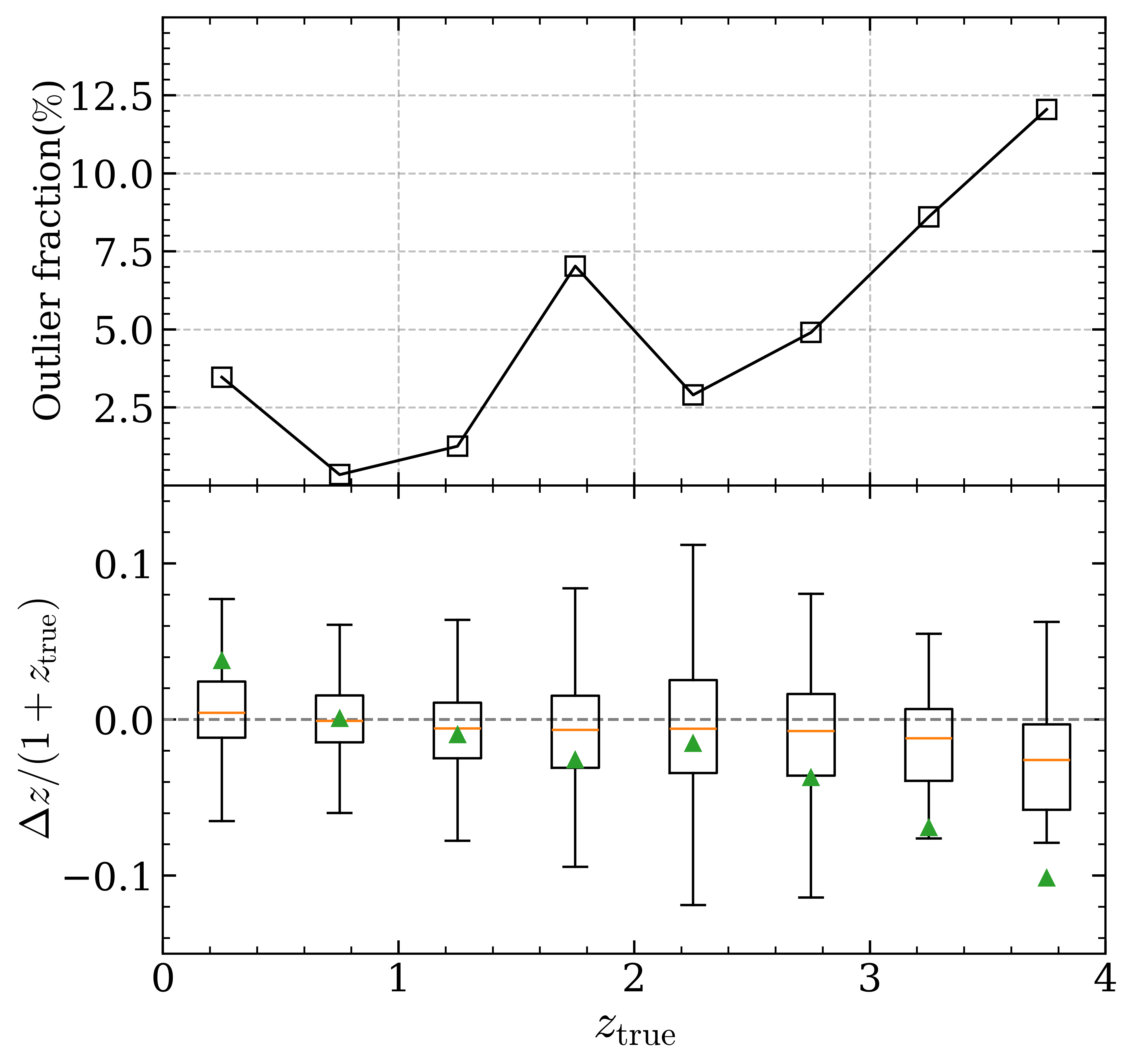}
	\caption{Upper panels are outlier fractions in different $z_{\rm true}$ bins. Bottom panels are box-plot of $\Delta z$ in the same bins. The triangle is the mean value of the whole data in current bins. The box in the plot represents the middle 50\% of the data, with the bottom edge of the box indicating the 25th percentile (Q1) and the top edge indicating the 75th percentile (Q3). The line inside the box represents the median (50th percentile). The whiskers extend from the box by 1.5x the inter-quartile range (IQR).}
	\label{fig:outliers}
\end{figure}

In Figure \ref{fig:outliers}, the top panel shows the outlier fraction in different redshift bins, while the bottom panel displays the box plot of $\Delta z /(1+z_{\rm true})$ in the same redshift bins. From the top panel, we can observe that in general, the outlier fraction increases with redshift. It tends to have the smallest fraction in the redshift range of $0.5<z_{\rm true}<1.5$, which is mainly due to the larger training samples in this range as shown in Figure \ref{fig:redshift}. In a larger redshift range of $3.5<z_{\rm true}<4$, there is an extreme increase in the outlier fraction, which could be mainlty caused by fewer training samples and larger observation flux errors. Note that there is a peak between $1.5<z_{\rm true}<2$, which is due to the observable wavelength limit of filter bands, as shown in Figure \ref{fig:filters}. Between $1.5<z_{\rm true}<2$, the Balmer break shifts out of the observation bands while Lyman break has just entered. As we can see, even without any physical model assumptions, the result still demonstrates a similar degeneracy problem in the template fitting method \citep{brammer2008eazy}. The box plot of $\Delta z/(1+z_{\rm true})$ below indicates that the mean (triangle) and median (middle line in box) prediction values decrease with redshift. The trend could be explained by the algorithm's inability to predict values beyond the training sample range. Thus, it is normal for the algorithm to overestimate at low redshifts and underestimate at high redshifts. 

\begin{figure}
	\centering
	\includegraphics[width=\columnwidth]{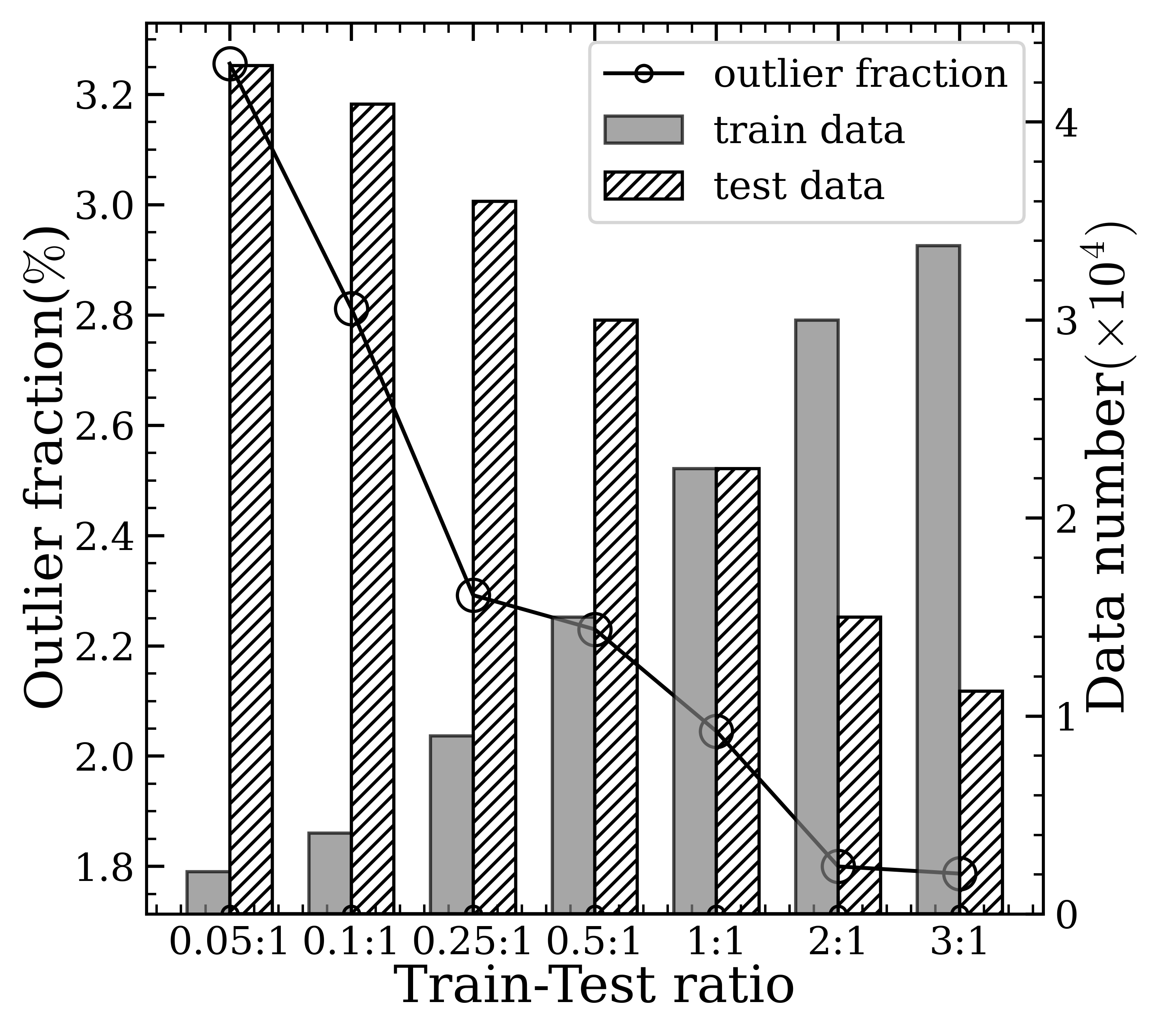}
	\caption{A comparison of different training sizes in RF. The solid line is the outlier fraction. Training and testing data numbers are shown by grey and shadow bars respectively.}
	\label{fig:ratios}
\end{figure}

The results presented above were obtained using a training-testing ratio of 1:1. To test if increasing this ratio would improve the photo-$z$ estimation, we tried different training-testing ratios as shown in Figure \ref{fig:ratios}. After applying the same process, we obtained the outlier fraction and $\rm \sigma_{NMAD}$ for the testing data. We observed that increasing the training data can lead to better predictions, although the improvement is very small. After the ratio exceeded 2:1, the improvements reached a limit. This suggests that the number of training data is sufficient, and increasing the ratio may not be meaningful. Moreover, the RF algorithm can still perform well even with a small amount of training data, such as $\sim0.25:1$. 

\subsection{Error distribution}
\begin{figure}
	\centering
	\includegraphics[width=\columnwidth]{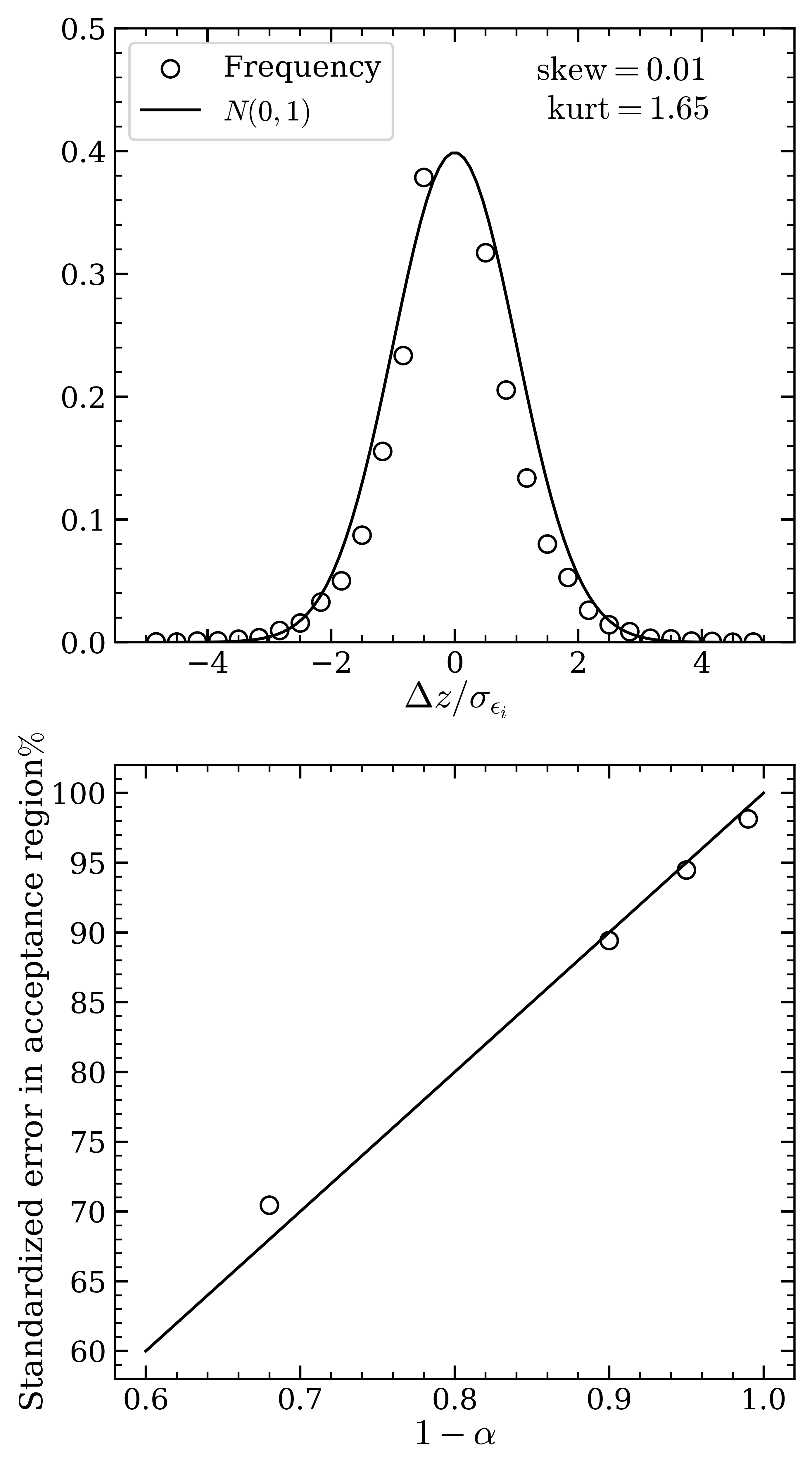}
	\caption{Upper panel shows a standard normal distribution (curve) and the distribution of standardized error $(z_{\rm phot}-z_{\rm true})/\sigma_{\epsilon_i}$ for the test data (circle). The lower panel shows the percent of error (circle) within level-$\alpha$ critical values. The straight line represents the percent error expected within level-$\alpha$. Horizontal axis is $1-\alpha$ \citep{carliles2010random}.}
	\label{fig:dztest}
\end{figure}

To examine the distribution of the difference between predicted and true values in our model, we calculated the standardized error $\Delta z/\sigma_{\epsilon_i}$ distribution, where $\sigma_{\epsilon_i}$ represents the standard deviation of each tree's prediction error for every source, which is presented as circles in the upper panel of Figure \ref{fig:dztest}. An unbiased model should yield a Gaussian distribution with a zero mean and a unit variance, represented by the curve on the plot. In general, the curve and histogram are well-matched, although there is a slight concentration near zero. Upon calculating the skew and kurtosis of the distribution, we found them to be 0.01 and 1.65, respectively, indicating that the data is almost symmetrical and unbiased, closely approximating a mesokurtic distribution.

We proceed to compute the percentage of standardized true errors that fall within the level-$\alpha$ critical values for a given $\alpha$ referenced from \citet{carliles2010random}. We plot several circles with $\alpha$ values of 0.32, 0.1, 0.05, and 0.01 to represent the values, and the line represents the expected percentage. For instance, if $\alpha=0.32$, the area under the lower or upper tail of a standard Gaussian distribution is $\alpha/2=0.16$. We expect the percentage to be the area between the two tails, which is $68\%$. We find that approximately $70.2\%$ of the error distribution falls within the critical values. While the fit is not perfect, the results are still close to the correct values within acceptable bounds.

\subsection{Probability distribution function}\label{sec:pdf-made}
\begin{figure*}
	\centering
	\includegraphics[width=\textwidth]{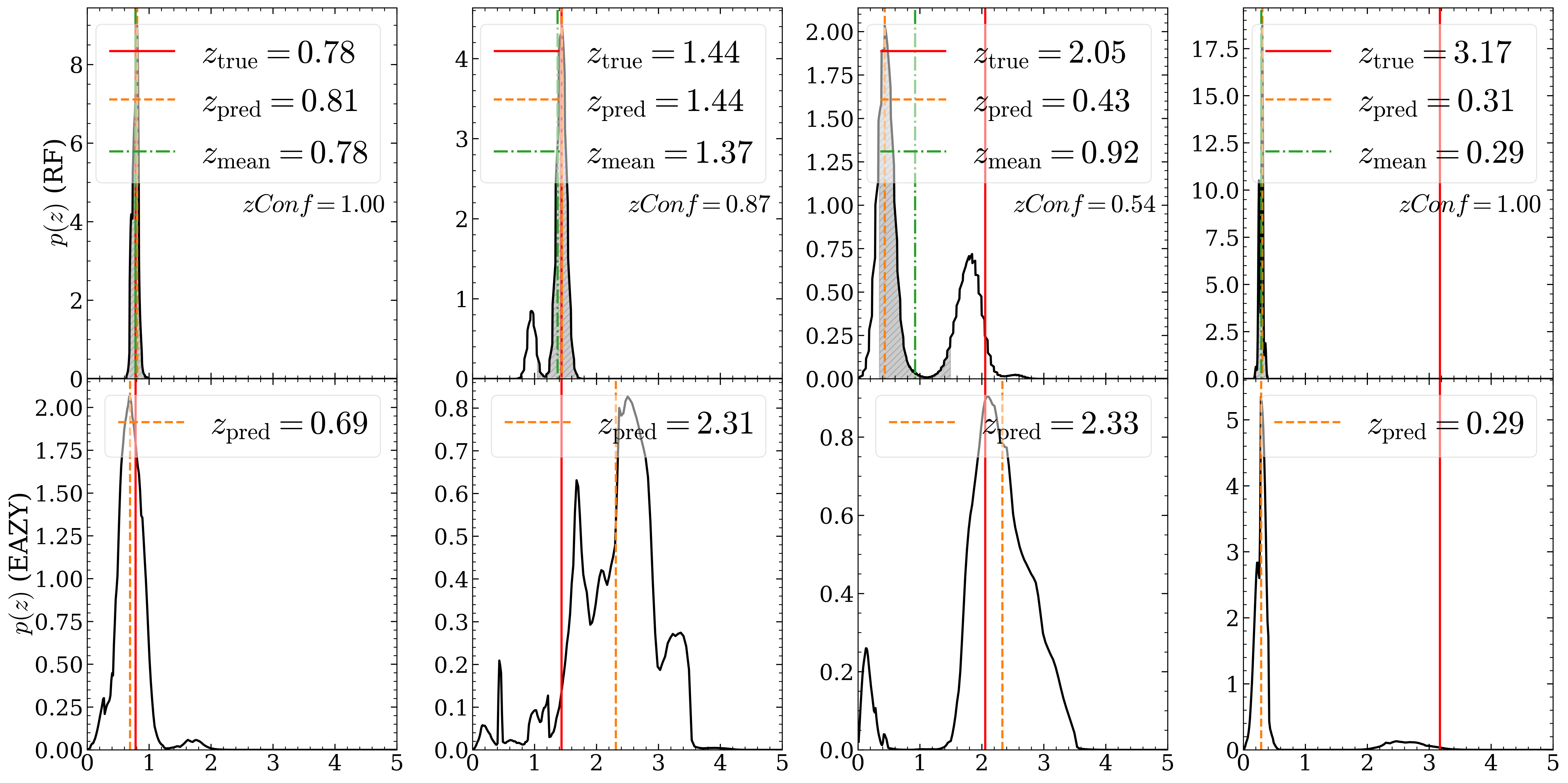}
	\caption{PDF examples obtained by RF (first row) and EAZY (second row). Every sub-figure in a column represents the same source, and the red solid line is true redshift. The dashed line is the predicted value. The average value is also shown by dot-dashed line. The grey area under the RF PDF is $zConf$. The larger the $zConf$, the more concentrated is around $z_{\rm mean}$.}
	\label{fig:pdf}
\end{figure*}

A classical PDF obtained by EAZY \citep{brammer2008eazy} is calculated from 
\begin{equation}\label{eq:eazypdf}
	p(z|m_0,C)\propto p(z|C)p(z|m_0),
\end{equation}
where $p(z|C)\propto {\rm exp}\left[-\chi^2/2\right]$ is the likelihood computed from the template fits over given redshift grids, $z$, and observed colours, $C$, $\chi^2$ is defined in Equation \ref{eq:chisquare}. $p(z|m_0)$ is the prior probability that a specific galaxy with magnitude $m_0$ at redshift $z$. In this work we construct the PDF made by EAZY using the prior calculated from $r$ band. The prior probability created by EAZY has the form of
\begin{equation}
	p(z|m_{0,i})\propto z^{\gamma_i}{\rm exp}\left[-(z/z_{0,i})^{\gamma_i}\right],
\end{equation}
where $\gamma_i$ and $z_{0,i}$ are the fitted parameters for the redshift distributions in each magnitude bin, $m_{0,i}$. Figure \ref{fig:pdf} displays some examples of PDF provided by our RF model (first row) and EAZY (second row). True and predicted redshifts are shown by solid and dashed lines. Here we notice that even though the PDF is concentrative, it still has the probability to predict inaccurately. 

It is important to clarify that this comparison doesn't imply the existence of a single superior model. In reality, there exists a multitude of methods for estimating photo-$z$ PDFs, yielding divergent outcomes, and a consensus on a preferred approach is yet to be established. Recent efforts have been made to evaluate the performance of various PDF estimation methods \citep{schmidt2020evaluation}.

Once the PDF is established, a new index called $zConf$ can be designed to represent the confidence level of the predicted redshift \citep{carrasco2013tpz}. $zConf$ is defined as the integral of the PDF within the range of $z_{\rm mean}\pm\alpha(1+z_{\rm mean})$, where $z_{\rm mean}$ is the mean value of the PDF. When $z_{\rm mean}<1$, $\alpha=0.3$, and when $z_{\rm mean}\geq1$, $\alpha=0.1\times(1+z_{\rm mean})$. Figure \ref{fig:pdf} shows the $zConf$ values for each PDF, where the grey area represents the value of $zConf$. In these examples, the $zConf$ around the mean value of each PDF is measured. Typically, the $zConf$ for galaxies with unimodal or concentrated PDFs is high, and the peak is more likely near the real redshift, while the $zConf$ for galaxies with bimodal or dispersed PDFs is low, indicating that $zConf$ can provide a reasonable confidence level for the redshift estimate. It should be noted that a low $zConf$ does not necessarily mean a poor prediction. Sometimes, the peak value of the PDF corresponds to the true redshift. Therefore, the index $zConf$ only represents the shape of the PDF rather than the accuracy. However, photo-$z$ samples can still be further refined by selecting only PDFs with $zConf$ higher than a certain threshold. Table \ref{tab:zconfcut} shows the proportion of galaxies retained after different $zConf$ cuts in RF and EAZY for comparison. The data depletion is smoother in the RF method, indicative of the RF algorithm generating a greater number of unimodal PDFs. Furthermore, the RF algorithm demonstrates superior accuracy across all cutoff conditions. Nonetheless, it is still possible that a small number of galaxies with high $zConf$ may be estimated as the wrong redshift, and similarly, a small number of sources with low $zConf$ may be estimated as the correct redshift.

\begin{table}
	\centering
	\caption{Different $zConf$ cut in testing data and EAZY.}
	\label{tab:zconfcut}
	\resizebox{\linewidth}{!}{
	\begin{tabular}{ccccc} 
		\hline
		 & $zConf$ cut & Data Remaining & Outlier Fraction & $\rm \sigma_{NMAD}$\\
		\hline
		RF &$zConf>0$    & 100\%  & 2.045\%  & 0.025\\
		& $zConf>0.6$  & 99.3\% & 1.714\%  & 0.025\\
		& $zConf>0.75$ & 99.1\% & 1.641\%  & 0.025\\
		 &$zConf>0.9$  & 98.5\% & 1.48\%   & 0.025\\
		\hline
		EAZY&$zConf>0$    & 100\%  & 6.45\%  & 0.043\\
		&$zConf>0.6$  & 91.5\% & 4.462\%  & 0.040\\
		&$zConf>0.75$ & 86.8\% & 3.77\%  & 0.039\\
		&$zConf>0.9$  & 78.6\% & 2.72\%   & 0.036\\
		\hline
	\end{tabular}
}
\end{table}

In addition to the confidence index, an error bar is added to the predicted value of each redshift. Based on the PDF of each galaxy, the redshift position of the 16th and 84th percentiles of the cumulative probability distribution is taken as the upper and lower limits of the error bar, as shown in Figure \ref{fig:confandbar}.
Like the confidence index, most outliers have large error bars. However, there are a small number of galaxies for which the error bar is very small, but the redshift is still incorrectly estimated. This situation may be due to insufficient input information.

\subsection{Feature importance}
In addition to the PDF, feature importance can provide helpful information. It helps us understand the training data better, determine the most effective feature combination, and test whether it is possible to reduce the number of input features. Diverse methods for feature importance assessment can yield varying results due to different measurement approaches, reliance on specific prediction-tailored features, and the impact of modeling techniques \citep{Saarela2021ComparisonOF}. Disparities in results may also emerge from the ability to capture mutual influences between features. Furthermore, using different data subsets or adjusting hyperparameters can introduce fluctuations in importance values. In our subsequent analysis, we conducted tests using various feature importance calculation methods.

\subsubsection{Model-dependent feature importance }
\begin{figure}
	\includegraphics[width=\columnwidth]{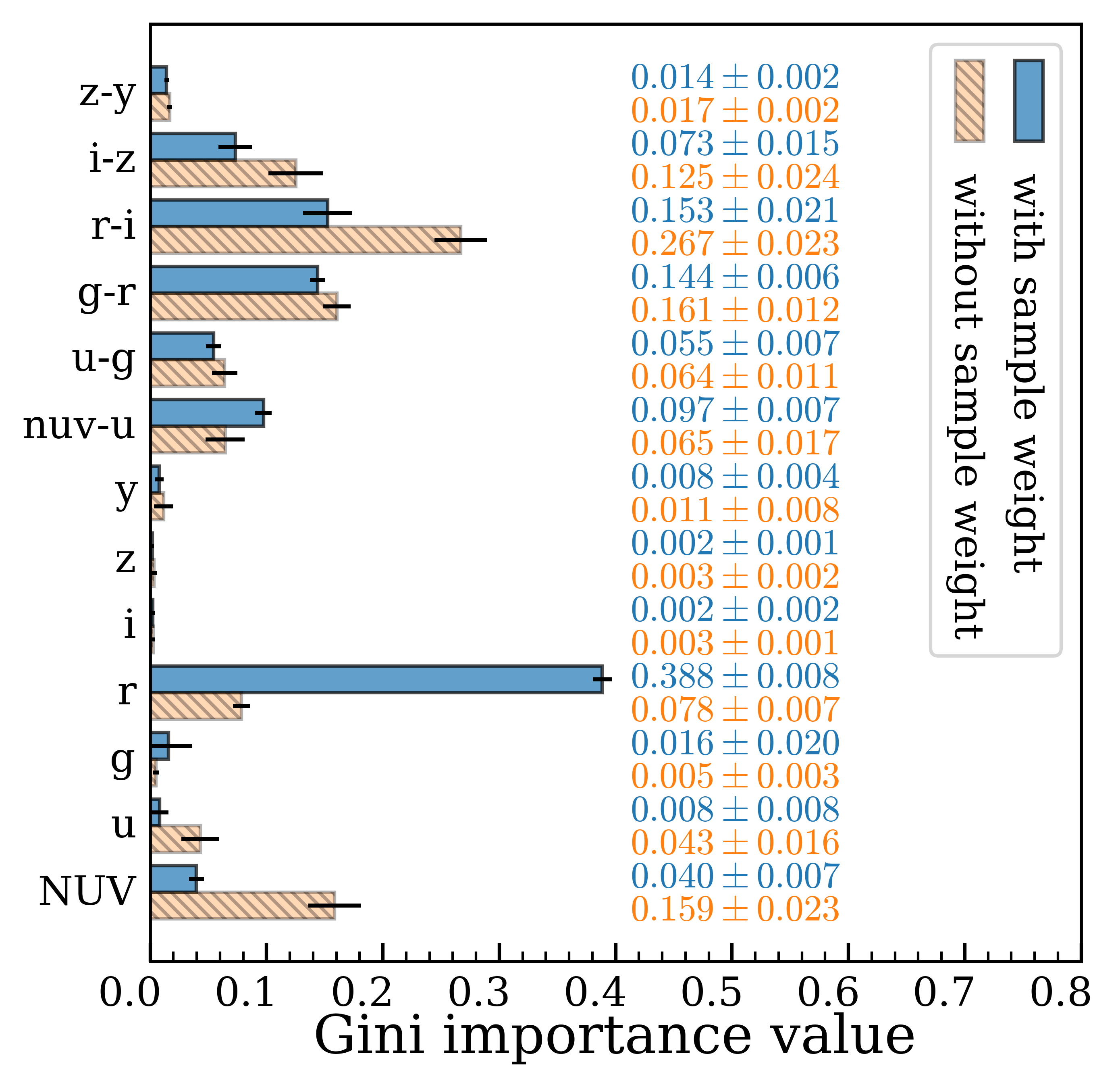}
	\includegraphics[width=\columnwidth]{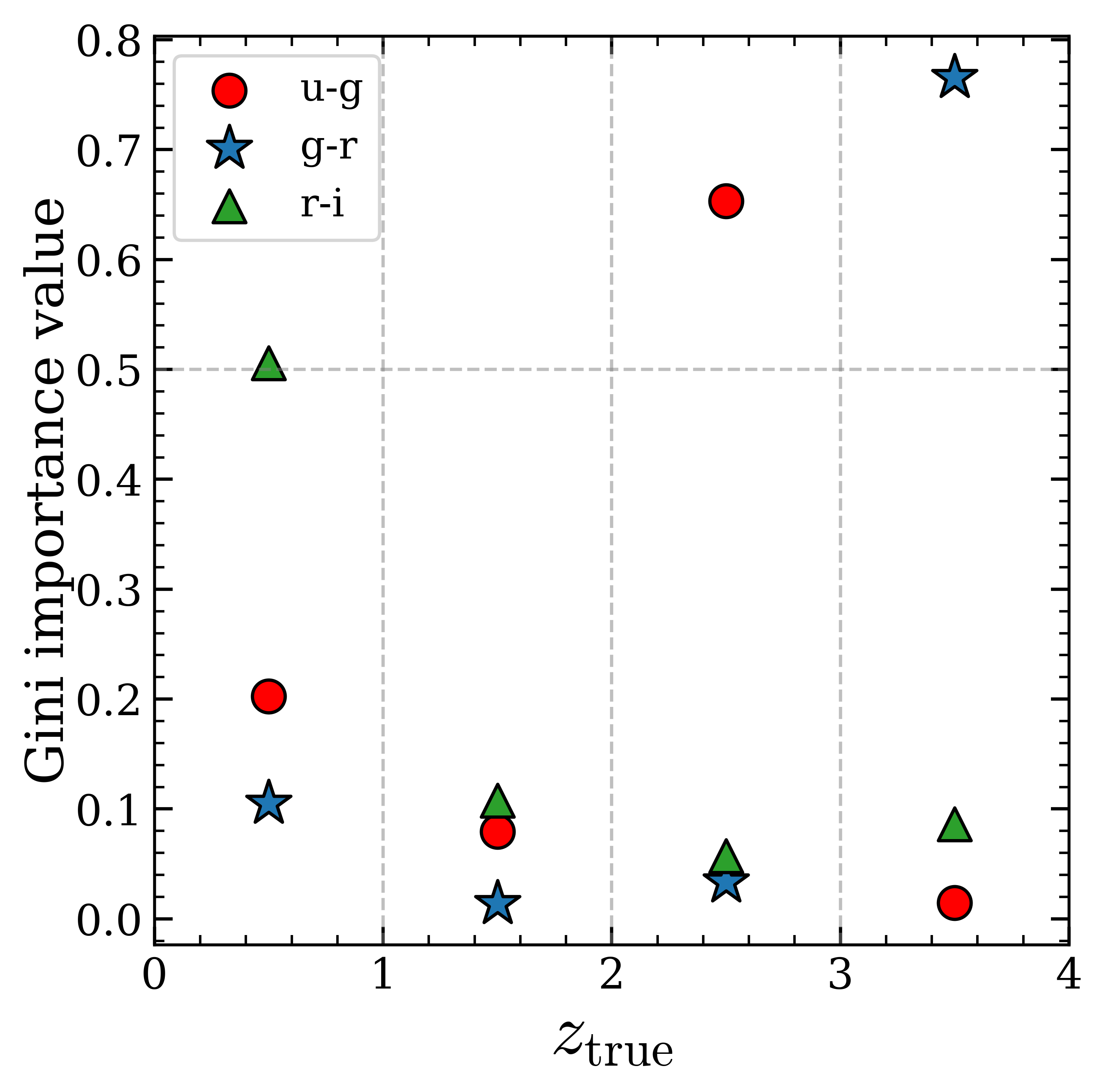}
	\caption{Feature importance of the training data. The upper panel is the importance of the whole data, The panel below is the importance of three features change with redshift.}
	\label{fig:importance}
\end{figure}

RF-based feature importance is a component of the output generated by RF models. These importances are calculated based on the mean and standard deviation of impurity decrease (as described in Equation \ref{eq:impuritydecrease}) accumulation within each tree, commonly referred to as Gini importance. Gini importance serves as a measure of the impact that each feature has on the predictive performance of the RF model. As shown in the upper panel of Figure \ref{fig:importance}, feature importance differs before and after adding sample weights to the model. This is expected because the weight parameter affects the random choice possibility of features when building a regression tree. The three most important features of the weighted model are the $r-i$, $g-r$, and $r$ bands. The reason why these two colors, along with the $r$ band luminosity, are important is because they can effectively constrain the shape of the SED for galaxies in our sample. The importance of the r-band is also reflected in EAZY code, which typically uses $r$ band data to construct the prior function. From the upper panel of Figure \ref{fig:importance} we can also see that any input features related to the $y$ band, such as $y$ band flux and $z-y$ color, do not exhibit significant importance in predicting photometric redshifts, primarily due to the fact that the wavelength range of the $y$ band is completely encompassed by the $z$ band. 

The input feature importance was further investigated by dividing the training samples into multiple bins based on redshift. The importance of $u-g$, $g-r$, and $r-i$  was then calculated for each bin. As shown in the lower panel of Figure \ref{fig:importance}, in the redshift range of $z\in(0,1]$, $r-i$ color is the most important as it reflects the Balmer (4000Å) break feature. In the $z\in(2,3]$ range, $u-g$ becomes the most important feature as the Balmer break has shifted out of the observed wavelength range and the Lyman break feature has entered the $u$ or $g$ band. Similarly, the $g-r$ color is most important in the $z\in(3,4]$ range for the same reason. Similar findings have also been reported in previous studies, indicating that the importance of different colors varies with redshift \citep{euclid2023selection,mucesh2021machine,li2023desi,zhou2019deep}. 

Adding more input features does not necessarily mean higher prediction accuracy. In fact, sometimes adding too many features can lead to overfitting and decrease the model's generalization ability. It is crucial to carefully choose the most relevant features and apply effective methods to process and filter them. To address this issue, we deleted the two least and most important features, respectively, and tested the performance of our model in each case. The results, summarized in table \ref{tab:importancecut}, showed that deleting the two least important features improved accuracy and sped up the running time of the code. This was because there were fewer random features when splitting nodes in the tree and less data to be saved in memory when building the regression tree, which improved the access speed of the cache. However, no matter how many features were deleted, deleting the most important feature had a significant impact on the results, as these features contain the most information needed to generate accurate photo-$z$. 

\begin{table}
	\centering
	\caption{A comparison in the accuracy of photo-$z$ prediction by using different feature combinations.}
	\label{tab:importancecut}
	\begin{tabular}{ccc} 
		\hline
		Feature Selection  & Outlier Fraction & $\rm \sigma_{NMAD}$\\
		\hline
		All features             &  2.045\%  & 0.025\\
		Remove 2 least important &  2.018\%  & 0.026\\
		Remove 2 most important  &  5.272\%  & 0.032\\
		\hline
	\end{tabular}
\end{table}

\subsubsection{Model-agnostic feature importance }

\begin{figure*}
	\centering
	\subfloat[]{\includegraphics[width=0.31\textwidth]{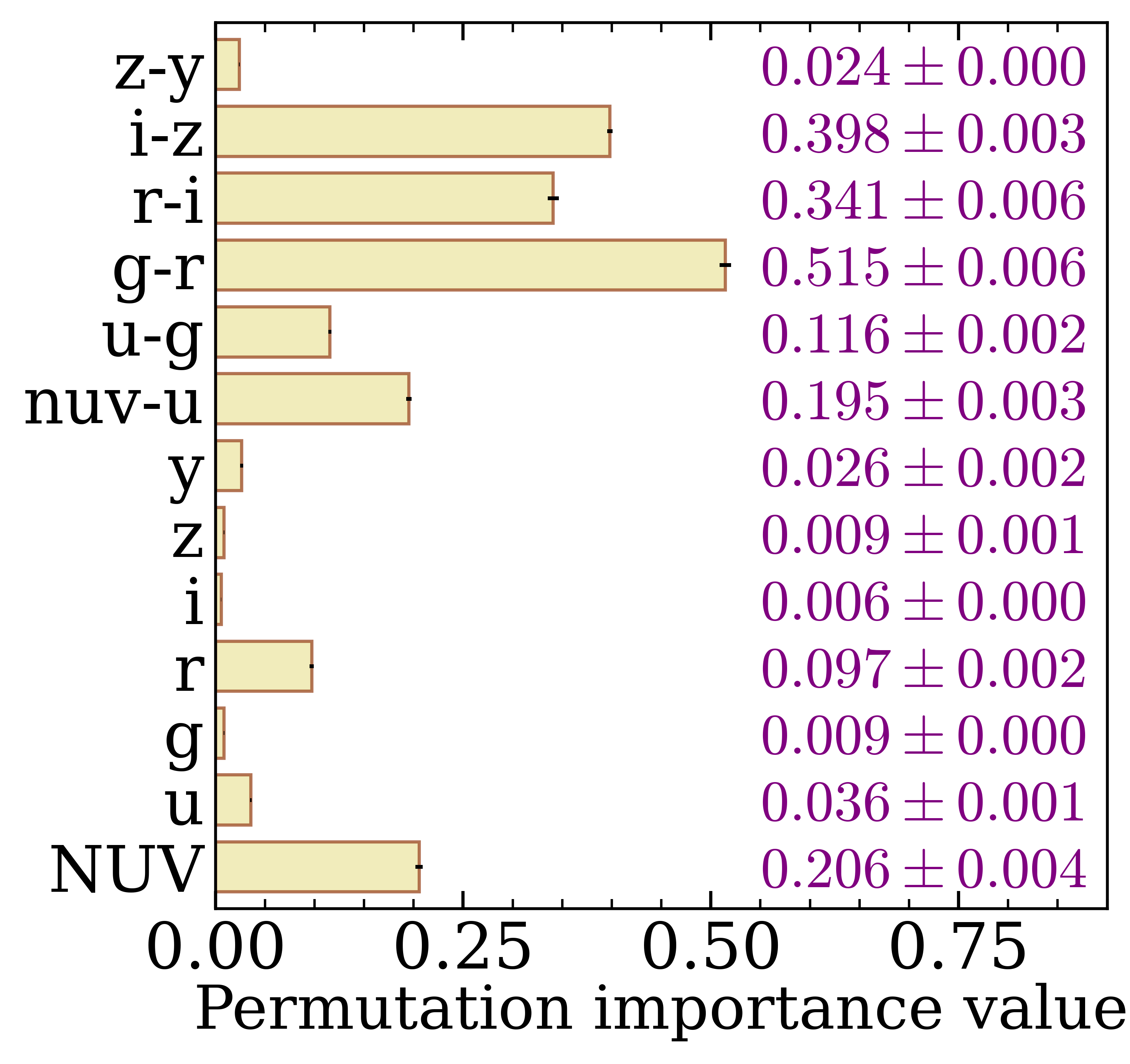}}
	\hfill
	\subfloat[]{\includegraphics[width=0.37\textwidth]{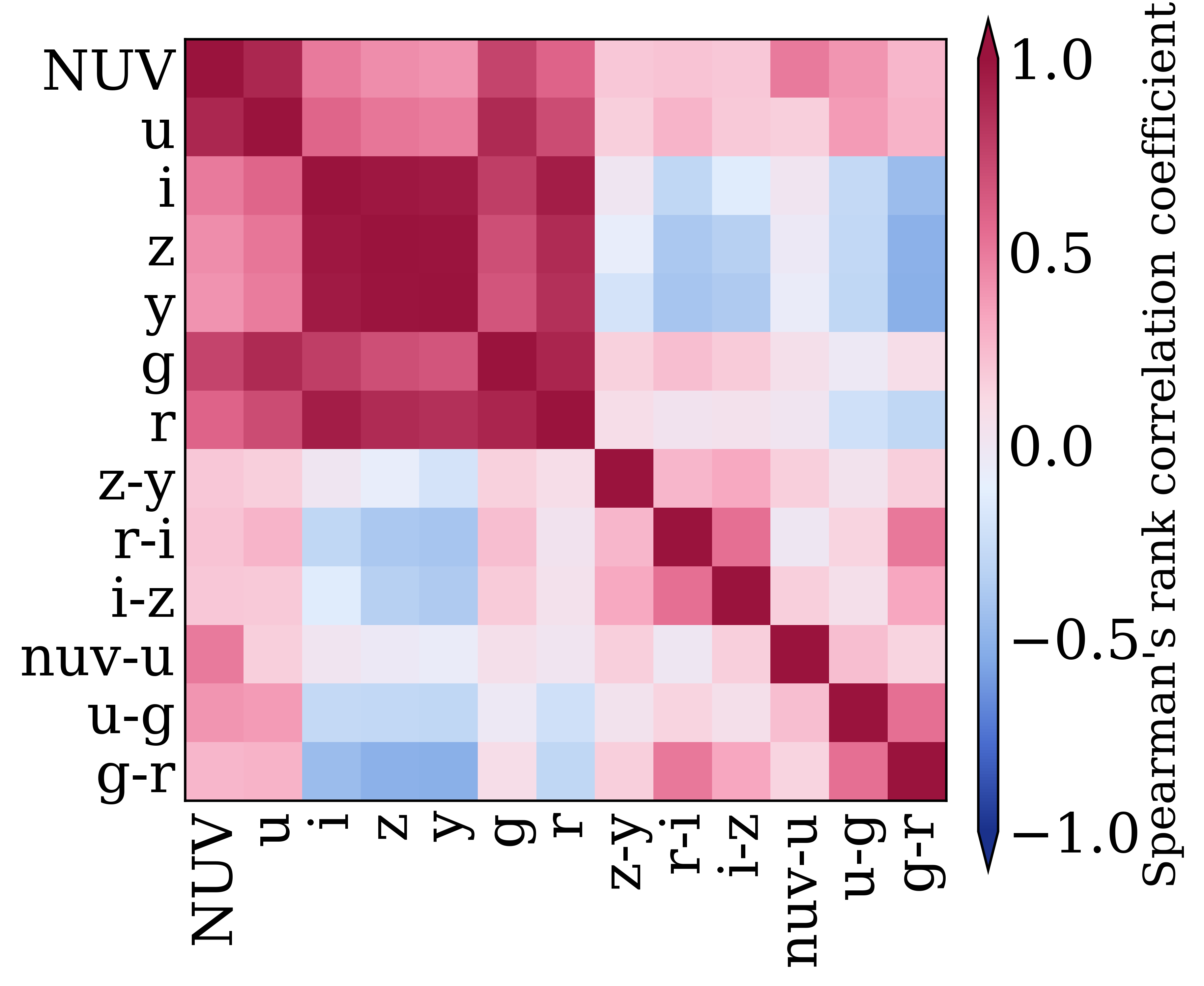}}
	\hfill
	\subfloat[]{\includegraphics[width=0.29\textwidth]{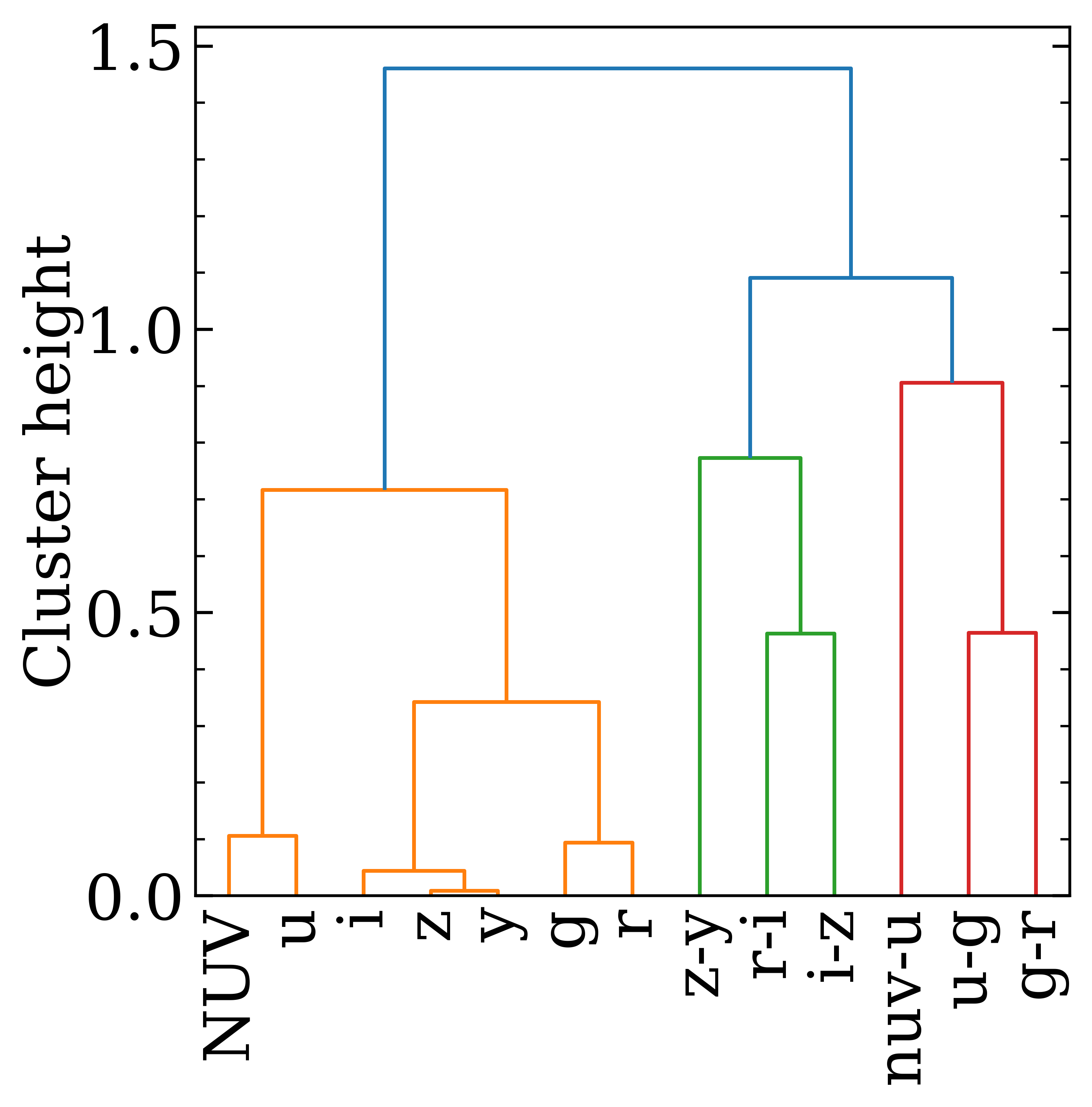}}	
	\caption{(a). Permutation importance of train set sample with sample weight. (b). Spearman's rank correlation coefficient of input features. (c). Hierarchical clustering dendrogram of input features with Ward's methods, the distance is calculated by $1-|\rm coefficient|$. The y-axis represents the similarity between features or clusters. Smaller values indicate higher correlation between two features or clusters.}
	\label{fig:permutation}
\end{figure*}
It is worth noting that when dealing with high cardinality features (those with numerous unique values) and in cases of significant feature co-linearity or correlation, impurity-based feature importances can potentially lead to misleading results, as discussed by \citep{euclid2023selection}. To examine the effect of feature co-linearity or redundant information among multiple features on importance values, we conducted several tests using different methods.

One of these methods is the permutation feature importance, as depicted in Figure \ref{fig:permutation}(a), where we computed the importance values for the input features in our training set. The permutation feature importance measures the reduction in a model's performance score when a single feature value is randomly shuffled. The importance value $i_j$ for feature $j$ is defined as follows: 
\begin{equation}
	i_j=s-\frac{1}{K}\sum_{k=1}^{K}s_{k,j}.
\end{equation} 
In this procedure, the variable $s$ represents the reference model score, and $s_{k,j}$ denotes the score for each repetition $k$ out of a total of $K$ data shuffles. By disrupting the relationship between the feature and the target variable, any decrease in the model score indicates the significance of the feature in the model's dependence. In situations where features are highly correlated, the permutation of one feature has limited impact on the model’s performance, as it can obtain similar information from a correlated feature. 

Based on the analysis from Figure \ref{fig:permutation}(a), it is evident that the top four important features, as identified through permutation feature importance, align closely with the Gini importance calculated in the RF model without sample weighting. Additionally, we can observe that both $i$ and $z$ are among the least important features. However, the combination feature $i-z$ carries greater importance than either of them individually.

Furthermore, we calculate the Spearman’s rank correlation coefficient \citep{zar2005spearman} between all features, as illustrated in Figure \ref{fig:permutation}(b). Then, we transform the correlation coefficient matrix into a distance matrix by computing $1-|\rm coefficient|$. This transformation aims to reduce the distance between features with higher correlation for subsequent hierarchical clustering. We employ Ward’s method \citep{ward1963Hierarchical} for clustering, and the clustering results are presented in Figure \ref{fig:permutation}(c).

Finally, we conducted tests on our model by selecting specific thresholds and retaining only one feature from each cluster. The performance of the model using these selected features was evaluated and is presented in Table \ref{tab:cluster screen}. This analysis provides insights into how the chosen features impact the performance of the model.
\begin{table}
	\centering
	\caption{The accuracy of our model for photometric redshift prediction by selecting specific thresholds and retaining only one feature from each cluster.}
	\label{tab:cluster screen}
	\begin{tabular}{cccc} 
		\hline
		Threshold & Feature Selection & Outlier Fraction & $\rm \sigma_{NMAD}$\\
		\hline
		1.2  & $NUV,g-r$     &  39.852\%  & 0.165\\
        \hline
		1.0  & $NUV,g-r,i-z$ &  21.688\%  & 0.066\\
        \hline
		0.8  & \makecell{$NUV,g-r,i-z,$\\$nuv-u$} &  15.585\%  & 0.053\\
        \hline
        0.6  & \makecell{$NUV,g-r,i-z,$\\$nuv-u,r,z-y$} &  7.881\%  & 0.037\\
        \hline
        0.4  & \makecell{$NUV,g-r,i-z,$\\$nuv-u,r,z-y,$\\$r-i,u-g$} &  2.005\%  & 0.025\\
		\hline
	\end{tabular}
\end{table}

\section{Summary}\label{section:conclusion}

In this work, we have used a weighted RF algorithm to predict the photo-$z$ of the CSST photometric survey. To simulate the CSST photometric observations, we use HST-ACS observations and the COSMOS catalog while taking into account the CSST instrumental effect. After obtaining the mock galaxy images in seven bands, fluxes and observation errors are measured using aperture photometry. We select sources with a signal-to-noise ratio greater than 10 in the $g$ or $i$ band as the training and testing sample for photo-$z$ prediction.

Our model takes the fluxes and colors of galaxies as input features and incorporates the inverse variance weight of the samples. During the training process, we also use Monte Carlo simulation to expand the training set by using input feature errors as standard deviations. The findings indicate that the weighted RF model can achieve a photo-$z$ accuracy of $\rm \sigma_{NMAD}=0.025$ and an outlier fraction of $\rm \eta=2.045\%,$ which is considerably better than the values of $\rm \sigma_{NMAD}=0.043$ and $\rm \eta=6.45\%$ obtained by the EAZY code that uses the template-fitting approach. Although our results are not superior to the prediction accuracy of \citet{Zhou_2022} based on a CNN, our model only requires the flux data of galaxies as input (colors can be derived from flux data) and does not need to input massive galaxy images like CNN, making the model prediction efficient. Additionally, the RF model can naturally provide the probability density distribution of predicted redshift and output the importance of each input feature. Therefore, our model can provide a methodological choice for effectively estimating photo-$z$ using CSST data.

Our research has found that the weighted RF model can accurately derive the photo-$z$ of a sample and can also establish confidence indices and error bars for each prediction value based on the derived redshift PDF. Selecting high-confidence sources can further reduce the proportion of outliers and improve prediction accuracy. Additionally, the model can determine the importance of each input feature in different redshift ranges. The results suggest that the most critical input feature typically reflects the approximate position of the spectral break in galaxies, indicating the model's ability to extract physical features. Moreover, our research also indicates that increasing the number of input features does not necessarily result in higher prediction accuracy. Only by optimizing the selection of input features can the prediction accuracy of the model be improved.

Further research has revealed that, despite the impressive results of the model, we are still unable to accurately measure the photo-$z$s of certain galaxies, regardless of how we adjust the model parameters or increase the number of training cycles. These galaxies are primarily concentrated in the low redshift range $z\sim(0-0.4)$ and high redshift range $z\sim(2-3)$. We believe that there are three reasons for this: firstly, there are relatively few training samples in these two redshift ranges, as can be seen from Figure \ref{fig:redshift}; secondly, the photometric accuracy of high redshift samples is relatively low due to lower flux, which often results in larger observational errors; thirdly, our model is unable to accurately distinguish between the Balmer break and Lyman break features in the galaxy's SED within the wavelength range observed by CSST, leading to overestimation of photo-$z$s in some low redshift galaxies and underestimation in some high redshift galaxies. Similar conclusions have also been mentioned in several previous studies\citep{euclid2023selection,mucesh2021machine,li2023desi,zhou2019deep}. Thus, the performance of our model is not solely dependent on the choice of model parameters and training cycles, but also on the quality of training samples and the range of wavelengths covered by observations.

\section*{Acknowledgements}
Z.J.L acknowledges the support from the Shanghai Science and Technology Fund under grant No. 20070502400, and the science research grants from the China Manned Space Project. L.P.F acknowledges the support from NSFC grant 11933002, and the Innovation Program 2019-01-07-00-02-E00032 of SMEC. WD acknowledges the support from NSFC grants No. 11890691. YG acknowledges the support from National Key R\&D Program of China grant Nos. 2022YFF0503404, 2020SKA0110402, the CAS Project for Young Scientists in Basic Research (No. YSBR-092), and China Manned Space Project with Grant Nos. CMS- CSST-2021-B01 and CMS-CSST-2021-A01. S.Z. is supported by The Program for Professor of Special Appointment (Eastern Scholar) at Shanghai Institutions of Higher Learning, the National Natural Science Foundation of China (Grant No. NSFC-12173026), and the Shanghai Committee of Science and Technology (Grant No. 23010503900 and 20ZR1473600). This work is also supported by the National Natural Science Foundation of China under Grants No. NSFC-12141302. ZF acknowledges the support from NSFC grant nos U1931210 and the China Manned Space Project with grant CMS-CSST-2021-A01. We would also like to extend our sincere gratitude to the anonymous reviewers for their meticulous review and valuable feedback on this manuscript. Their thoughtful comments and suggestions have greatly enriched the quality and depth of this work. We are truly appreciative of the time and expertise they dedicated to this review process.

\section*{DATA AVAILABILITY}
The data that support the findings of this study are available from the corresponding author, upon reasonable request.




\bibliographystyle{mnras}
\bibliography{export} 




\appendix

\section{Random Forest}\label{app:rf}
RF is made up of prediction trees, which can be divided into two categories: regression trees and classification trees \citep{breiman1984friedman}. They share similar calculation principles. For our work, we use the regression tree to solve the prediction problem. Figure \ref{fig:regrtree} shows a typical regression tree. The tree recursively partitions the training set into a hierarchy of clusters with similar objects by asking a series of questions until reaching the stop condition, such as the minimum number of samples in the leaf node. The subsample contained in a node represents a specific feature in the entire data set. The prediction tree generated by the model provides an intuitive way to understand the prediction process when many variables may interact in a nonlinear manner.

A regression tree is typically a binary tree, meaning a node is divided into two branch nodes. There are two main differences between regression trees and classification trees. Firstly, each leaf node of the regression tree contains training data with different redshift values, and the predicted value is given based on these data points, usually the average value. Therefore, the prediction result is no longer a discrete classification but an estimation of continuous variables. Secondly, the best split point algorithm for regression trees is based on impurity. In the regression tree, impurity is often calculated by minimizing the sum of the mean square deviation in each node. For node $R$, the sum of mean square deviation MSD is given by the following formula:
\begin{figure*}
	\includegraphics[width=\textwidth]{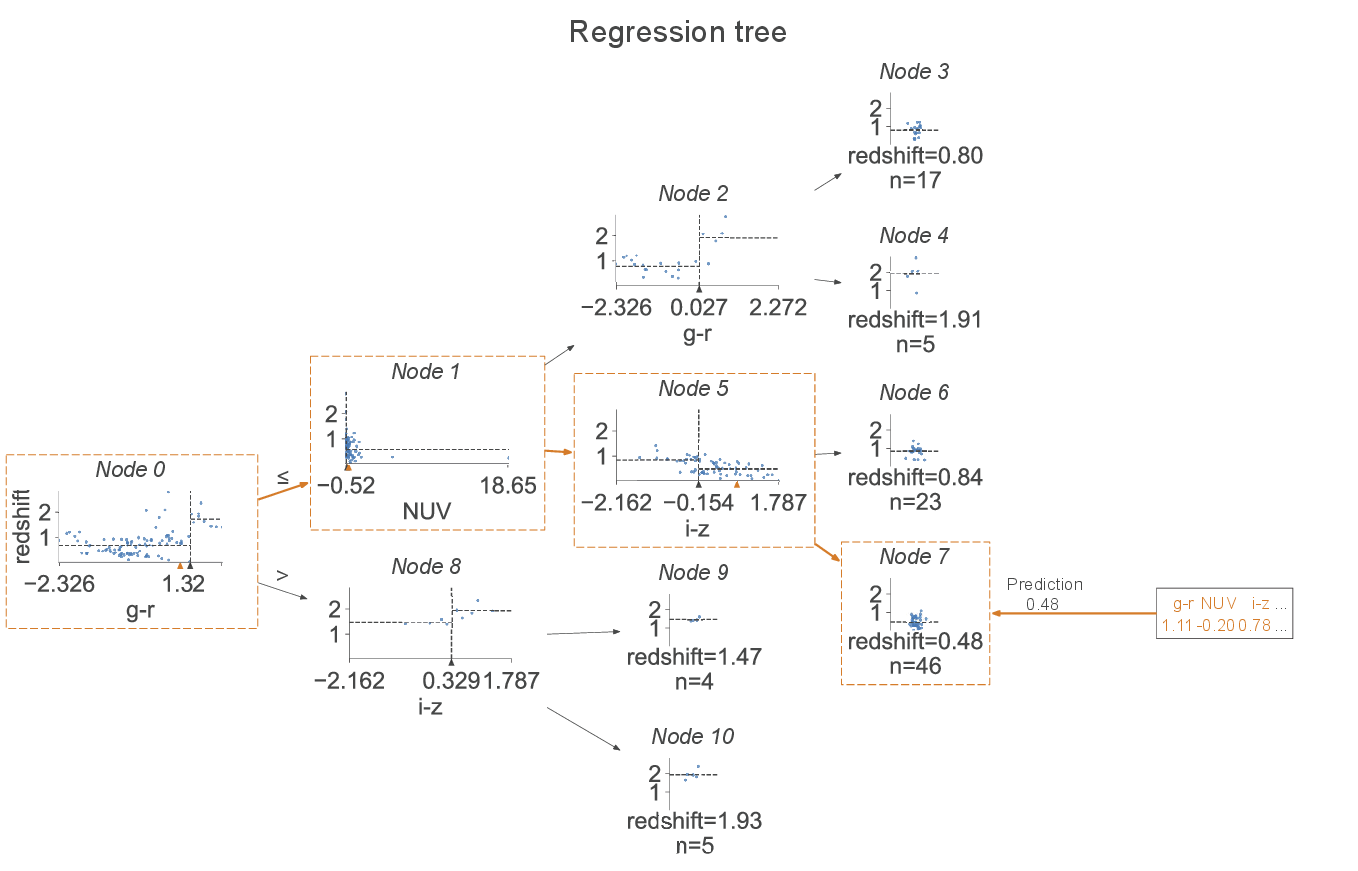}
	\caption{An example of a simple regression tree. The root node is on the left. The algorithm will split data in each node until reaches the stop criterion. The horizontal axis in each node is the best split feature. Each leaf node provides the final prediction based on the choice process.}
	\label{fig:regrtree}
\end{figure*}

\begin{equation}
	{\rm MSD} = \min\limits_{j,s}\left[\min\limits_{c_1}\sum_{x_i\in R_1(j,s)}(y_i-c_1)^2+\min\limits_{c_2}\sum_{x_i\in R_2(j,s)}(y_i-c_2)^2\right],
	\label{eq:MSE}
\end{equation}
That is, for any given split feature $j$, the corresponding split point $s$ divides the data into two sets $R_1$ and $R_2$. The algorithm then searches for the optimal feature and split point that minimizes the mean square deviation in each node, while ensuring that the sum of both sets is minimized. In this formula, $c_1$ and $c_2$ represent the mean values in sets $R_1$ and $R_2$, respectively. The binary tree algorithm will recursively apply this process on each new node until a stopping criterion, such as reaching a minimum number of samples in the node, is met. 

The basic idea of RF is to create a large number of prediction trees and then combine their predictions to improve accuracy. Given a training sample $R$ with $N$ objects and $M$ features (e.g., flux, color, size), the RF algorithm creates $N_T$ subsamples of size $N$ by randomly selecting data points from the parent sample with replacement. This process is known as bootstrapping. For each subsample, a corresponding prediction tree is created. The estimates from these trees are then averaged to obtain a more robust prediction. In addition to providing accurate estimates, RF can also provide a distribution of individual estimates.

If all variables are considered when determining the best splitting point, the method is called bagging \citep{breiman1996bagging}. RF can incorporate additional randomization by selecting the best input features from a random subspace $m_*$ instead of selecting the best feature from the entire input space $M$. For example, a node may only consider three features out of $u$, $g$, $r$, and $i$, such as $u$, $r$, and $z$. The value of $m_*$ is an adjustable parameter that can increase the independence between trees and reduce computational cost. The more correlated the trees are, the higher the error rate of RF. The final accuracy of the algorithm is not very sensitive to a large number of trees and a relatively small number of input features. RF combines all prediction estimates to achieve a more robust prediction.

During the training process, it is possible to assign weights to individual samples to specify their importance when building the regression trees. In scikit-learn, the sample weights can be specified through the "sample\_weight" parameter, which takes an array of weights assigned to each training sample. The algorithm adjusts the weights of each tree in the forest to reflect the importance of each sample during training. This is achieved by using the weighted sampling technique, where samples with higher weights have a higher probability of being selected for inclusion in the tree, while samples with lower weights have a lower probability.

When splitting a node in a regression tree, the algorithm calculates the impurity of each possible split and selects the split that minimizes the impurity. If sample weights are provided, the impurity of each possible split is weighted by the sample weights. The weighted impurity decrease equation is given as follows:

\begin{equation}
	\begin{aligned}
		{\rm impurity~decrease} &= \frac{N_t}{N}*({\rm impurity} \\
		&-\frac{N_{t_R}}{N_t}*{\rm right\_impurity}\\
		& -\frac{N_{t_L}}{N_t}*{\rm left\_impurity}),
	\end{aligned}
	\label{eq:impuritydecrease}
\end{equation}
where variables $N$, $N_t$, $N_{t_R}$, and $N_{t_L}$ in regression tree algorithms refer to the total number of samples, the number of samples at the current node, the number of samples in the right child, and the number of samples in the left child, respectively. When the parameter "sample\_weight" is passed, these values are computed as weighted sums. This means that samples with higher weights contribute more to the calculation of impurity, and the algorithm will prioritize splits that result in lower impurity for the most important samples. The use of sample weights can be beneficial in situations where some samples are more important than others or when dealing with imbalanced datasets where the model needs to be trained to give more weight to the minority class. However, it is crucial to use sample weights with care and thoughtfulness as they can introduce bias into the model if not used appropriately.



\bsp	
\label{lastpage}
\end{document}